\documentclass[12pt]{article}
\usepackage{amsmath,amssymb,latexsym,epsf,epsfig,graphicx,tikz}
\pdfoutput=1

\begin{document}

\newcommand{\sect}[1]{\setcounter{equation}{0}\section{#1}}
\renewcommand{\theequation}{\thesection.\arabic{equation}}
\newcommand{\prt}{\partial}
\newcommand{\II}{\mbox{${\mathbb I}$}}
\newcommand{\CC}{\mbox{${\mathbb C}$}}
\newcommand{\RR}{\mbox{${\mathbb R}$}}
\newcommand{\QQ}{\mbox{${\mathbb Q}$}}
\newcommand{\ZZ}{\mbox{${\mathbb Z}$}}
\newcommand{\NN}{\mbox{${\mathbb N}$}}
\def\G{\mathbb G}
\def\UU{\mathbb U}
\def\S{\mathbb S}
\def\tS{\widetilde{\mathbb S}}
\def\V{\mathbb V}
\def\tV{\widetilde{\mathbb V}}
\newcommand{\D}{{\cal D}}
\def\hint{H_{\rm int}}

\newcommand{\rd}{{\rm d}}
\newcommand{\diag}{{\rm diag}}
\newcommand{\U}{{\cal U}}
\newcommand{\cP}{{\cal P}}

\newcommand{\ph}{\varphi}
\newcommand{\phd}{\widetilde{\varphi}} 
\newcommand{\phs}{\varphi^{(s)}}
\newcommand{\phb}{\varphi^{(b)}}
\newcommand{\phds}{\widetilde{\varphi}^{(s)}}
\newcommand{\phdb}{\widetilde{\varphi}^{(b)}}
\newcommand{\lambdad}{\widetilde{\lambda}}
\newcommand{\tx}{\widetilde{x}} 
\newcommand{\phl}{\varphi_{i,L}}
\newcommand{\phr}{\varphi_{i,R}}
\newcommand{\phz}{\varphi_{i,Z}}
\newcommand{\mur}{\mu_{{}_R}}
\newcommand{\mul}{\mu_{{}_L}}
\newcommand{\muv}{\mu_{{}_V}}
\newcommand{\mua}{\mu_{{}_A}}

\def\a{\alpha}

\def\A{\mathcal A} 
\def\C{\mathcal C} 
\def\O{\mathcal O}
\def\I{\mathcal I}
\def\der{\partial }
\def\mis{{\frac{\rd k}{2\pi} }}
\def\ri{{\rm i}}
\def\xt{{\widetilde x}}
\def\ft{{\widetilde f}}
\def\gt{{\widetilde g}}
\def\qt{{\widetilde q}}
\def\tt{{\widetilde t}}
\def\tmu{{\widetilde \mu}}
\def\prt{{\partial}}
\def\tr{{\rm Tr}}
\def\inc{{\rm in}}
\def\out{{\rm out}}
\def\e{{\rm e}}
\def\eps{\varepsilon}
\def\k{\kappa}
\def\v{{\bf v}}
\def\ebf{{\bf e}}
\def\abf{{\bf A}}


\newcommand{\finprf}{\null \hfill {\rule{5pt}{5pt}}\\[2.1ex]\indent}

\pagestyle{empty}
\rightline{June 2011}

\vfill

\begin{center}
{\Large\bf Non-equilibrium Steady States of\\ Quantum Systems on Star Graphs}
\\[2.1em]

\bigskip

{\large Mihail Mintchev}\\ 

\null

\noindent 

{\it  
Istituto Nazionale di Fisica Nucleare and Dipartimento di Fisica, Universit\`a di
Pisa, Largo Pontecorvo 3, 56127 Pisa, Italy}
\vfill

\end{center}
\begin{abstract}

Non-equilibrium steady states of quantum fields on star graphs are explicitly constructed. 
These states are parametrized by the temperature and the chemical potential, associated 
with each edge of the graph. Time reversal invariance is spontaneously broken. 
We study in this general framework the transport properties of the Schr\"odinger and the 
Dirac systems on a star graph, modeling a quantum wire junction.
The interaction, which drives the system away from equilibrium, is 
localized in the vertex of the graph. All point-like vertex interactions, 
giving rise to self-adjoint Hamiltonians possibly involving the minimal coupling to a static 
electromagnetic field in the ambient space, are considered. In this context we compute the exact 
electric steady current and the non-equilibrium charge density. We investigate also the heat transport and derive the 
Casimir energy density away from equilibrium. The appearance of Friedel type oscillations of the 
charge and energy densities along the edges of the graph is established. We focus finally on 
the noise power and discuss the non-trivial impact of the point-like interactions on the noise.

\end{abstract}
\bigskip 
\medskip 
\bigskip 

\vfill
\rightline{IFUP-TH 10/2011}
\newpage
\pagestyle{plain}
\setcounter{page}{1}

\section{Introduction} 
\medskip 

\begin{figure}[h]
\begin{center}
\begin{picture}(550,150)(80,305) 
\includegraphics[scale=1]{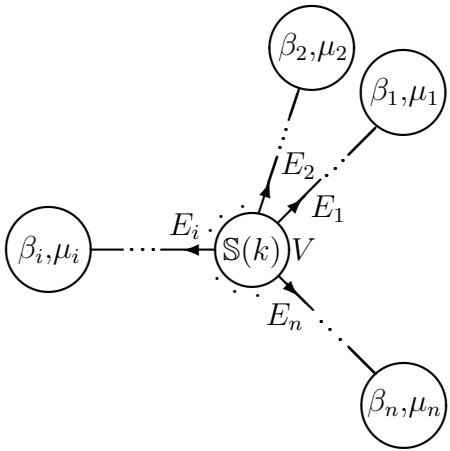}
\end{picture} 
\end{center}
\caption{A star graph $\Gamma$ with scattering matrix $\S(k)$ at the vertex $V$ 
and edges $E_i$ connected at infinity to thermal reservoirs with inverse temperature 
$\beta_i$ and chemical potential $\mu_i$.} 
\label{junction}
\end{figure}

Transport properties of quantum wire junctions attracted in the last two decades much attention \cite{kf-92}-\cite{Soori:2010ga}. 
The experimental realizations of quantum wires include nowadays carbon nanotubes, semiconductor, 
metallic and polymer nanowires, and quantum Hall edges. While the equilibrium features of 
these devises has been extensively explored, there is recently a growing interest in 
non-equilibrium phenomena. A typical problem in this context is schematically represented 
in Fig. \ref{junction}. A quantum wire junction is modeled by a star graph $\Gamma$ with $n$  
edges (leads) $E_i$, each of them connected to a heat reservoir (bath) with (inverse) temperature 
$\beta_i$ and chemical potential $\mu_i$. Assume for simplicity that $E_i$ are infinite and that the 
interaction, described by a scattering matrix $\S(k)$, is localized at the vertex $V$ of $\Gamma$.  
The system is away from equilibrium if $\S(k)$ admits at least one non-trivial transmission coefficient 
among edges with different temperature and/or chemical potential. Two main questions arise at this point. 
The first one concerns the existence of a steady state 
describing the above situation. Provided that such a state exists, it is natural to ask about the general features  
of the system in this state. In the present paper we show that the first question 
has an affirmative answer, constructing explicitly a suitable steady state $\Omega_{\beta, \mu}$, 
which is parametrized by $\beta = (\beta_1,...,\beta_n)$ and $\mu = (\mu_1,...,\mu_2)$ and captures 
the non-equilibrium properties of the system. Afterwards we derive the expectation 
values of several basic  physical observables (currents, charge and energy densities,...) in $\Omega_{\beta, \mu}$, which 
characterize the system and thus answer the second question. 
\bigskip 

It is perhaps useful to recall that the systems admitting non-equilibrium steady states (NESS's) \cite{mcl-59} 
represent an important subclass of the large family of non-equilibrium systems. Unlike in equilibrium, 
a system in a NESS admits steady currents. Nevertheless, all macroscopic properties are still time 
independent like in equilibrium. 

The formulation of a suitable statistical mechanical framework for 
treating NESS's has not been yet completed and is intensively investigated \cite{els-96}-\cite{d-10}. 
We follow below a microscopic approach and address the question using quantum field theory methods. 
Our construction of NESS generalizes that of a Gibbs state in finite temperature quantum field theory. We 
develop first a general algebraic framework, which applies for any dispersion relation,  
thus covering both the relativistic and non-relativistic cases. 
As already mentioned, we consider for simplicity fields which propagate freely 
in the bulk $\Gamma \setminus V$ of the graph, the interaction being localized exclusively in the 
vertex $V$. We discuss all possible point-like interactions for which the Hamiltonian of our system, 
being a Hermitian operator in the bulk $\Gamma \setminus V$, extends to a self-adjoint operator on the 
whole graph $\Gamma$. We use at this point of our construction some simple results 
\cite{ks-00}-\cite{k-08} from the spectral theory of differential operators on graphs (known also as ``quantum graphs"). 

It is worth mentioning that non-equilibrium systems of the type shown in Fig. \ref{junction} 
have been investigated in the past by two different methods. The first one is the scattering approach 
initiated by Landauer \cite{la-57} and completed by B\"uttiker \cite{bu-86}. The second one is based on 
the linear response theory \cite{KTH}, 
originally developed by Kubo for macroscopic samples and extended 
later by Baranger and Stone \cite{bs-89} to mesoscopic systems. The equivalence between these two 
approaches has been demonstrated in Ref. \cite{cj-05}. 

The construction of the NESS $\Omega_{\beta, \mu}$, developed in this paper, 
is formulated in purely algebraic terms and adopts 
a deformation of the algebra of canonical (anti)commutation relations. No approximations, like linear 
response theory, are involved. The states $\Omega_{\beta, \mu}$ allow for a unified description of systems with 
different dynamics. We illustrate this fact by treating below both the Schr\"odinger and Dirac equations in 
the same way. The abstract construction of $\Omega_{\beta, \mu}$ is first tested by reproducing 
the Landauer-B\"uttiker (L-B) steady current. Afterwards, taking advantage of the exact solvability of the 
point-like interactions in the state $\Omega_{\beta, \mu}$, we investigate several new features of the 
non-equilibrium dynamics, applying the general framework to other physical observables. We 
compute the charge density in $\Omega_{\beta, \mu}$ and show the presence of characteristic Friedel 
type oscillations along the leads. We also present a fully microscopic-based calculation of the Casimir 
effect in a quantum wire junction away from equilibrium. In this context we establish the non-equilibrium 
Stefan-Boltzmann law and derive the heat (energy) transport. We also compute the exact 
two-point current-current correlation function away from equilibrium and investigate the noise power. 
In particular, we discuss the impact of the point-like interaction in the junction on the behavior of 
both thermal and shot noises. Finally, we generalize all results in the presence of a static electromagnetic field 
in the ambient space. 

The paper is organized as follows. In the next section we develop a simple algebraic framework for the 
the construction of steady states of quantum wire junctions. Section 3 is devoted to the Schr\"odinger 
junction. In section 4 we change the dynamics, considering the Dirac equation and the relative 
steady states on the junction. The novelty here is the presence of antiparticles which contribution and 
role are considered in detail. In section 5 we discuss the results and comment on the possible 
further developments. 

\bigskip

\section{Algebraic construction of the states $\Omega_{\beta, \mu}$} 
\medskip 

Previous investigations \cite{Bellazzini:2006jb}-\cite{Bellazzini:2008fu}, \cite{Sch}  
have shown that a convenient coordinate system for 
expressing quantum fields on star graphs is provided by the deformed algebras 
$\A_\pm$ of canonical (anti)commutation relations, generated by 
$\{a_i(k),\, a^*_i(k)\, :\, k \in \RR, \, i=1,...,n\}$ which satisfy 
\begin{equation} 
[a_i(k)\, ,\, a_j(p)]_\pm = [a^*_i (k)\, ,\, a^*_j (p)]_\pm = 0 \, , 
\label{rta1}
\end{equation}
\begin{equation}
[a_i(k)\, ,\, a^*_j (p)]_\pm = 2\pi [\delta (k-p)\delta_{ij} + \S_{ij}(k)\delta(k+p)] \, , 
\label{rta2}
\end{equation}  
and the constraints 
\begin{equation} 
a_i(k) = \sum_{j=1}^n \S_{ij} (k) a_j (-k) \, , \qquad 
a^\ast_i (k) = \sum_{j=1}^n a^\ast_ j(-k) \S_{ji} (-k)\, .    
\label{constr1}
\end{equation} 
The index $\pm$ in (\ref{rta1}, \ref{rta2}) refers to Fermi/Bose statistics and 
$\S(k)$ is the $n\times n$ scattering matrix 
describing the point-like interaction in the vertex of the graph. 
We assume in what follows unitarity 
\begin{equation}
\S(k) \S(k)^* = \II 
\label{unit}
\end{equation} 
and Hermitian analyticity \cite{Liguori:1996xr}-\cite{Mintchev:2004jy}  
\begin{equation}
\S(k)^*=\S(-k) \, . 
\label{ha}
\end{equation} 
The latter implies that the $*$-operation is a conjugation in $\A_\pm$. Combining (\ref{unit}) and (\ref{ha}) one 
concludes that $\S(k) \S(-k) = \II$, ensuring the consistency of the constraints (\ref{constr1}). 

Our main goal now is to construct $\Omega_{\beta, \mu}$ as a state, i.e. as a 
(continuous) linear functional over $\A_\pm$. We recall in this respect that 
$\A_\pm$ is a simplified version of the so called 
reflection-transmission algebra \cite{Liguori:1996xr}-\cite{Mintchev:2004jy}, 
describing factorized scattering in integrable models with point-like defects in one dimension. 
The Fock and the (grand canonical) Gibbs state over $\A_\pm$ describe equilibrium physics 
and have been largely explored \cite{Liguori:1996xr}-\cite{Mintchev:2004jy}. 
The physical input for constructing the new states $\Omega_{\beta, \mu}$ is the 
observation that the sub-algebras $\A_\pm^\inc$ and $\A_\pm^\out$, generated by $\{a_i(k),\, a^*_i(k)\, :\, k<0\}$ 
and $\{a_i(k),\, a^*_i(k)\, :\, k>0\}$ respectively, parametrize the asymptotic incoming and outgoing fields. 
Accordingly, both $\A_\pm^\inc$ and $\A_\pm^\out$ are conventional canonical 
(anti)commutation relation algebras. Indeed, the $\delta (k+p)$ term in (\ref{rta2}) vanishes if 
both momenta are negative or positive. Notice also that the constraints (\ref{constr1}) relate 
$\A_\pm^\inc$ with $\A_\pm^\out$. It is crucial for what follows that the whole reflection-transmission 
algebra $\A_\pm$ can be generated either by $\A_\pm^\inc$ or by $\A_\pm^\out$ via (\ref{constr1}). 
Our strategy for constructing the NESS $\Omega_{\beta , \mu}$ is based on this kind of asymptotic completeness 
property. In fact, we will start with an equilibrium state on $\A_\pm^\inc$ and extend it by means of 
(\ref{constr1}) to a non-equilibrium state on $\A_\pm$. 

The first step is to describe the {\it asymptotic} dynamics and symmetries at $t=-\infty$ 
(i.e. before the interaction) in terms of $\A_\pm^\inc$. Since the asymptotic fields are free, it is 
natural to introduce the edge Hamiltonians 
\begin{equation} 
h_i = \int_{-\infty}^0 \frac{\rd k}{2\pi} \omega_i (k) a^*_i(k) a_i(k) 
\label{ah}
\end{equation} 
and edge charges
\begin{equation} 
q_i = \int_{-\infty}^0 \frac{\rd k}{2\pi} a^*_i(k) a_i(k)\, ,   
\label{ac}
\end{equation} 
where $\omega_i(k)\geq 0$ is the dispersion relation in the edge $E_i$. At this point we define 
\begin{equation} 
K = \sum_{i=1}^n \beta_i (h_i -\mu_i q_i) \, . 
\label{kop}
\end{equation} 
and introduce the equilibrium Gibbs state over $\A_\pm^\inc$ in the standard way \cite{BR}. For 
any polynomial $\cP$ over $\A_\pm^\inc$ we set 
\begin{equation}
\left (\Omega_{\beta , \mu}\, ,\,  \cP(a_i^*(k_i), a_j(p_j)) \Omega_{\beta , \mu} \right ) \equiv 
\langle \cP(a_i^*(k_i), a_j(p_j)) \rangle_{\beta, \mu} = 
\frac{1}{Z} \tr \left [\e^{-K} \cP(a_i^*(k_i), a_j(p_j))\right ]\, ,  
\label{def1}
\end{equation} 
where $k_i<0,\; p_j<0$ and $Z = \tr \left ( \e^{-K}\right )$. 
It is well known \cite{BR} that one can compute the expectation values (\ref{def1}) 
by purely algebraic manipulations and that all these expectation values can be expressed in 
terms of the two-point functions, which are written in terms of 
the familiar Fermi/Bose distributions in the following way 
\begin{equation}
\langle a_j^*(p)a_i(k) \rangle_{\beta, \mu} = 
\frac{\e^{-\beta_i [\omega_i (k) - \mu_i]}}{1\pm \e^{-\beta_i [\omega_i (k) - \mu_i]}} 
\delta_{ij} 2\pi \delta (k-p)\, ,  
\label{2a}
\end{equation}
\begin{equation} 
\langle a_i(k)a_j^*(p)\rangle_{\beta, \mu} = 
\frac{1}{1\pm \e^{-\beta_i [\omega_i (k) - \mu_i]}} 
\delta_{ij} 2\pi \delta (k-p)\, .  
\label{2b}
\end{equation}
We stress that (\ref{2a},\ref{2b}) hold on $\A_\pm^\inc$, i.e. only for {\it negative} momenta. 

The second step is to extend (\ref{def1}-\ref{2b}) to 
the whole algebra $\A_\pm$, namely to {\it positive} momenta. For this 
purpose we use the relations (\ref{constr1}). One finds in this way  
\begin{eqnarray} 
\langle a_j^*(p)a_i(k)\rangle_{\beta, \mu} = 
2\pi \Bigl\{\Bigl [\theta(-k)d^\pm_i(k) \delta_{ij}+ 
\theta(k)\sum_{l=1}^n \S_{il}(k)\, d^\pm_l(-k)\, \S_{lj}(-k)\Bigr ] \delta (k-p)  
\nonumber \\
+ \Bigl [\theta(-k)d^\pm_i(k) \S_{ij}(k) + \theta(k)\S_{ij}(k) d^\pm_j(-k) \Bigr ] \delta (k+p) \Bigr\}\, ,
\qquad \; \; \,  
\label{cor1}
\end{eqnarray} 
were for simplifying the notation we introduced  
\begin{equation} 
d^\pm_i (k) = \frac{\e^{-\beta_i [\omega_i (k) - \mu_i]}}{1\pm \e^{-\beta_i [\omega_i (k) - \mu_i]}} \, .
\label{fbd1} 
\end{equation}
The explicit expression of $\langle a_i(k)a_j^*(p)\rangle_{\beta, \mu}$ is 
obtained from (\ref{cor1}) by the substitution 
\begin{equation} 
d^\pm_i(k) \longmapsto c^\pm_i(k) =  
 \frac{1}{1\pm \e^{-\beta_i [\omega_i (k) - \mu_i]}} \, . 
\label{fbd2} 
\end{equation} 

The final step is to compute a generic correlation function. Employing the commutation 
relations (\ref{rta1},\ref{rta2}), one can reduce the problem to the evaluation of correlators of the form 
\begin{equation}  
\langle \prod_{m=1}^M a_{i_m}(k_{i_m}) \prod_{n=1}^N a^\ast_{j_n}(p_{j_n})\rangle_{\beta,\mu} \, ,  
\label{gcf1}
\end{equation}
which can be computed in turn by iteration via 
\begin{equation} 
\langle \prod_{m=1}^M a_{i_m}(k_{i_m}) \prod_{n=1}^N a^{\ast j_n}(p_{j_n})
\rangle_{\beta,\mu} = 
\delta_{MN}\, \sum_{m=1}^M \langle a_{i_1}(k_{i_1})a^{\ast j_m}(p_{j_m})\rangle_{\beta,\mu}  
\, \langle \prod_{m=2}^M a_{i_m}(k_{i_m}) 
\prod_{n=1\atop {n\not=m} }^N a^{\ast j_n}(p_{j_n}) \rangle_{\beta,\mu} \, .   
\label{gcf2}
\end{equation} 

In conclusion, we emphasize once more that the use of the deformed algebras $\A_\pm$ in the construction of 
$\Omega_{\beta, \mu}$ represents only a convenient choice of coordinates, which has 
the advantage to be universal and to apply to a variety of systems characterized by a 
scattering matrix $\S(k)$. In support of this statement, we consider below 
the Schr\"odinger and the Dirac equations on the star graph $\Gamma$. 
\bigskip 

\section{The Schr\"odinger junction}

\subsection{Preliminaries} 
\medskip 

In this section we apply the general algebraic construction of the state $\Omega_{\beta, \mu}$ to the 
Schr\"odinger system on a star graph $\Gamma$ with point-like interactions in the vertex $V$ of $\Gamma$. 
We will consider Fermi statistics and for simplifying the notation will omit the apex $+$ in the Dirac distributions 
$d^+_i(k)$ and $c^+_i(k)$. As observed in section 3.5 below, most of the results can be easily extended to 
Bose statistics. 

We start by summarizing the main features \cite{Bellazzini:2006jb} of the Schr\"odinger equation on $\Gamma$, 
recalling the description of all point-like interactions leading to a self-adjoint Hamiltonian. 
Each point $P$ in the bulk of $\Gamma$ is parametrized by $(x,i)$, where $x>0$ is the distance 
of $P$ from the vertex $V$ and $i$ labels the edge. In the bulk $\Gamma \setminus V$ of the graph 
the Schr\"odinger field $\psi(t,x,i)$ with Fermi statistics satisfies  
\begin{equation}
\left (\ri \prt_t +\frac{1}{2m} \prt_x^2\right )\psi (t,x,i) = 0\, , 
\label{eqm1}
\end{equation}
with standard equal-time 
canonical anticommutation relations 
\begin{equation}
[\psi (0,x_1,i_1)\, ,\, \psi (0,x_2,i_2)]_+ = 
[\psi^* (0,x_1,i_1)\, ,\, \psi^* (0,x_2,i_2)]_+=0  \, , 
\label{initial1} 
\end{equation}
\begin{equation}
[\psi (0,x_1,i_1)\, ,\, \psi^* (0,x_2,i_2)]_+ =  
\delta_{i_1i_2}\, \delta (x_1-x_2)  \, .  
\label{initial2}
\end{equation} 
The interaction in the vertex is fixed by requiring that the bulk Hamiltonian defined by (\ref{eqm1}) (essentially the 
operator $-\prt_x^2$) admits a {\it self-adjoint extension} on the whole graph. According to some 
elementary results from the spectral theory \cite{ks-00}-\cite{k-08} of differential operators on graphs, 
all such interactions are described by the boundary conditions 
\begin{equation} 
\lim_{x\to 0^+}\sum_{j=1}^n \left [\lambda (\II-\UU)_{ij} -\ri (\II+\UU)_{ij}\prt_x \right ] \psi (t,x,j) = 0\, , 
\label{bc1} 
\end{equation} 
where $\UU$ is an arbitrary $n\times n$ unitary matrix and $\lambda \in \RR$ is a 
parameter with dimension of mass. Eq. (\ref{bc1}) guaranties {\it unitary time evolution} of the 
system and generalizes to the graph $\Gamma$ the familiar mixed (Robin) boundary condition on the half-line $\RR_+$. The matrices $\UU=\II$ and $\UU=-\II$ define the Neumann and Dirichlet 
boundary conditions respectively. 

The explicit form of the scattering matrix, expressed in terms of $\UU$ and $\lambda$, is \cite{ks-00}-\cite{k-08} 
\begin{equation} 
\S (k) = -\frac{[\lambda (\II - \UU) - k(\II+\UU )]}{[\lambda (\II - \UU) + k(\II+\UU )]}   
\label{S1}
\end{equation} 
and has a simple physical interpretation: the diagonal element $\S_{ii}(k)$ 
represents the reflection amplitude from the vertex on the edge $E_i$, whereas  
$\S_{ij}(k)$ with $i\not=j$ equals the transmission amplitude from $E_i$ to $E_j$. 
One easily verifies that (\ref{S1}) satisfies (\ref{unit}, \ref{ha}) and therefore defines an algebra 
$\A_+$ of the type introduced in the previous section. Notice that 
\begin{equation} 
\S(\lambda ) = \UU\, , \quad \S(-\lambda ) = \UU^{-1} \, , 
\label{S2} 
\end{equation} 
showing that the unitary matrix $\UU$ entering the boundary conditions (\ref{bc1}) is actually the scattering 
matrix at scale $\lambda$. 

A remarkable property of (\ref{S1}) is that it can be diagonalized for any $k$ by a 
$k$-independent unitary matrix. In fact, let $\U$ be the unitary matrix diagonalizing $\UU$, namely  
\begin{equation} 
\U^{-1}\, \UU\, \U = \UU_d=  \diag \left (\e^{2\ri \alpha_1}, \e^{2\ri \alpha_2}, ... , \e^{2\ri
\alpha_n}\right )\, , \qquad \alpha_i \in \RR\, . 
\label{d1}
\end{equation}  
By means of (\ref{S1}) one concludes that $\U$ diagonalizes $\S(k)$ {\it for any} $k$ as well, and that 
\begin{equation} 
\S_d(k) = \U^{*} \S(k) \U = \\
\diag \left (\frac{k+\ri \eta_1}{k-\ri \eta_1}, \frac{k+\ri \eta_2}{k-\ri \eta_2}, ... , \frac{k+\ri \eta_n}{k-\ri \eta_n} \right ) \, , 
\label{d3}
\end{equation} 
where 
\begin{equation} 
\eta_i = \lambda \tan (\alpha_i)\, , 
\qquad -\frac{\pi}{2} \leq \alpha_i \leq \frac{\pi}{2}\, .  
\label{d4}
\end{equation} 
Illustrating various aspects of the Schr\"odinger junction, 
we will use below the most general $2\times 2$ $\S$-matrix 
\begin{equation} 
\S(k)= \left(\begin{array}{cc}\frac{k^2 + \ri k (\eta_1-\eta_2)\cos(\theta)+\eta_1 \eta_2}{(k-\ri \eta_1)(k-\ri \eta_2)} 
& \frac{-\ri \e^{\ri \varphi} k (\eta_1-\eta_2)\sin(\theta)}{(k-\ri \eta_1)(k-\ri \eta_2)}\\ 
\frac{-\ri \e^{-\ri \varphi} k (\eta_1-\eta_2)\sin(\theta)}{(k-\ri \eta_1)(k-\ri \eta_2)} 
& \frac{k^2 - \ri k (\eta_1-\eta_2)\cos(\theta)+\eta_1 \eta_2}{(k-\ri \eta_1)(k-\ri \eta_2)}  \\ \end{array} \right)\, ,   
\label{NN0}
\end{equation}
where $\varphi$ and $\theta$ are arbitrary angles. 

The general representation (\ref{d3}) implies that $\S(k)$ is a meromorphic function 
in the complex $k$-plane with finite number of simple poles on the imaginary axis. For 
simplicity we consider in this paper the case without bound states (poles in the upper half plane), 
referring for the general case to \cite{Mintchev:2004jy}, \cite{Bellazzini:2008cs}, \cite{Bellazzini:2010gs} 
and the comments in section 3.5. In other words, we assume that 
\begin{equation}
\int_{-\infty}^{\infty} \frac{dk}{2\pi }\, \e^{\ri kx}\, \S_{ij}(k) = 0\, , \quad x>0\, . 
\label{compl1}
\end{equation}  
It has been shown in Ref. \cite{Bellazzini:2006jb} that in this case the solution of equation (\ref{eqm1}) is fixed 
uniquely by (\ref{initial1}-\ref{bc1}) and takes the following simple form
\begin{equation}
\psi (t,x,i) = \int_{-\infty}^{\infty} 
\frac{dk}{2\pi }
a_i (k) \e^{-\ri \omega (k)t+\ri kx}\, , \qquad  \omega(k) = \frac {k^2}{2m} \, . 
\label{psi1} 
\end{equation}
The equation of motion (\ref{eqm1}) is invariant under the time reversal operation
\begin{equation}
T \psi (t,x,i) T^{-1} = -\eta_T \psi (-t,x,i)\, , \qquad  |\eta_T|=1\, ,
\label{tr1}
\end{equation}
$T$ being an anti-unitary operator. We stress however that the boundary condition (\ref{bc1}) 
preserves the time reversal symmetry only if $\UU$ is symmetric \cite{Bellazzini:2009nk}, namely
\begin{equation}
\UU^t = \UU\, . 
\label{tr2}
\end{equation}

The electric current 
\begin{equation}
j_x(t,x,i)= \ri \frac{e}{2m} \left [ \psi^*(\partial_x\psi ) - (\partial_x\psi^*)\psi \right ]  (t,x,i) \, , 
\label{curr1}
\end{equation}
$e$ being the electric charge, and the energy current  
\begin{equation}
\theta_{xt} (t,x,i) = \frac{1}{4m}  [\left (\partial_t \psi^* \right )\left (\partial_x \psi \right ) 
+ \left (\partial_x \psi^* \right )\left (\partial_t \psi \right ) \\ - 
\left (\partial_t \partial_x \psi^* \right ) \psi - 
\psi^*\left (\partial_t \partial_x \psi \right ) ](t,x,i) \, , 
\label{en1} 
\end{equation} 
are among the basic physical observables. The time components of these currents are 
\begin{equation}
j_t (t,x,i)=  e \left ( \psi^*\psi \right )  (t,x,i)\, ,   
\label{dens1}
\end{equation} 
\begin{equation}
\theta_{tt} (t,x,i) = -\frac{1}{4m}  \left[ \psi^* \left (\partial_x^2 \psi \right )+
\left (\partial_x^2 \psi^* \right )\psi \right] (t,x,i) \, , 
\label{endens1} 
\end{equation}
respectively and, as a consequence of (\ref{eqm1}), satisfy the local conservation laws 
\begin{equation} 
\left (\partial_t j_t - \partial_x j_x \right )(t,x,i) = 
\left (\partial_t \theta_{tt} - \partial_x \theta_{xt} \right )(t,x,i) = 0 \, .
\label{cons1} 
\end{equation} 
Equations (\ref{cons1}), combined with the Kirchhoff rules 
\begin{equation} 
\sum_{i=1}^n j_x(t,0,i) = 0\, , \qquad 
\sum_{i=1}^n \theta_{xt}(t,0,i) = 0 \, , 
\label{KK1}
\end{equation}
ensure the charge and energy conservation in the system. Since the proof of (\ref{KK1}) at the quantum level is a 
quite subtle, we provide the main steps, focussing for instance on the electric current. 
Using the basic definitions, one easily derives the representation
\begin{eqnarray}
j_x(t, 0,i) = \ri \frac{e}{2m} \int_{-\infty}^0 \frac{\rd k}{2\pi}  \int_{-\infty}^0 \frac{\rd p}{2\pi} 
\e^{\ri t  [\omega(k) - \omega(p)]} 
\qquad \qquad \quad \nonumber \\
\times \sum_{j, l=1}^n a^\ast_j(k) 
\Bigl \{ \chi^*_{ji}(k;0) \left [\der_{x} \chi_{il}\right ](p;0) - \left [\der_{x} \chi^*_{ji}\right ](k;0) \chi_{il}(p;0) \Bigr \} a_l(p) \, , 
\label{KK2}
\end{eqnarray}
where 
\begin{equation}
\chi(k;x) = \e^{\ri k x} \II + \e^{-\ri kx}\, \S^*(k)\, , \qquad 
\chi^\ast (k;x) = \e^{-\ri k x} \II + \e^{\ri kx}\, \S(k) \, .
\label{KK3}
\end{equation}
\bigskip 
The trick now is represent the right hand side of (\ref{KK3}) as a boundary term of an integral over the half line, namely 
\begin{eqnarray}
j_x(t, 0,i) = -\ri \frac{e}{2m} \int_{-\infty}^0 \frac{\rd k}{2\pi}  
\int_{-\infty}^0 \frac{\rd p}{2\pi} \int_0^\infty \rd x \, \e^{\ri t  [\omega(k) - \omega(p)]} 
\qquad \qquad \qquad \nonumber \\
\times \sum_{j, l=1}^n a^\ast_j(k) 
\Bigl \{ \chi^*_{ji}(k;x) \left [\der^2_{x} \chi_{il}\right ](p;x) - 
\left [\der^2_{x} \chi^*_{ji}\right ](k;x) \chi_{il}(p;x) \Bigr \} a_l(p) = 
\qquad \qquad \nonumber \\ 
-\ri \frac{e}{2m} \int_{-\infty}^0 \frac{\rd k}{2\pi}  \int_{-\infty}^0 \frac{\rd p}{2\pi} 
\e^{\ri t  [\omega(k) - \omega(p)]} (k^2-p^2)  \int_0^\infty \rd x \sum_{j, l=1}^n a^\ast_j(k) 
\left [\chi^*_{ji}(k;x) \chi_{il}(p;x) \right ]a_l(p) \, , 
\label{KK4}
\end{eqnarray} 
The final step is to apply the orthogonality relations 
\begin{equation} 
\sum_{i=1}^n \int_0^\infty \rd x\, \chi^*_{ji}(k;x) \chi_{il}(p;x)  = 2\pi \delta_{jl} \delta (k-p) \, , 
\label{KK5} 
\end{equation}
which hold \cite{Bellazzini:2006jb} for $\S$ given by (\ref{S1}). 

At this stage we are ready to investigate the properties of the system in the state $\Omega_{\beta, \mu}$. 
Being expressed in terms of the algebra $\A_+$, the solution (\ref{psi1}) and the observables 
(\ref{curr1}-\ref{endens1}) apply for any representation of this algebra. 
This fundamental universality property has been already largely explored in the Fock 
and the Gibbs representations of $\A_+$, which describe equilibrium physics. In order to study the non-equilibrium 
properties of the Schr\"odinger system in Fig. \ref{junction}, we apply below the representation generated 
by the state $\Omega_{\beta, \mu}$. Since antiparticle excitations are absent, we assume without loss of generality that 
$\mu_i \geq 0$. 

There are two non-trivial two-point correlation functions. Using (\ref{cor1}), one finds 
\begin{eqnarray}
\langle \psi^*(t_1,x_1,i) \psi(t_2,x_2,j)\rangle_{\beta, \mu} = 
\int_0^{\infty} \frac{\rd k}{2\pi} \e^{\ri \omega(k) t_{12}} 
\qquad \qquad \qquad \qquad \qquad 
\nonumber \\
\left [\delta_{ji} d_i(k) \e^{-\ri k x_{12}} + 
\S_{ji}(k) d_j(k) \e^{\ri k \tx_{12}} + \S^\ast_{ji}(k) d_i(k) \e^{-\ri k \tx_{12}} + 
\sum_{l=1}^n \S^\ast_{jl}(k) d_l(k) \S_{li}(k) \e^{\ri k x_{12}} \right ] , 
\label{corr11}
\end{eqnarray}
\begin{eqnarray}
\langle \psi (t_1,x_1,i) \psi^\ast (t_2,x_2,j)\rangle_{\beta, \mu} = 
\int_0^{\infty} \frac{\rd k}{2\pi} \e^{-\ri \omega(k) t_{12}} 
\qquad \qquad \qquad \qquad \qquad 
\nonumber \\
\left [\delta_{ij} c_i(k) \e^{\ri k x_{12}} + 
\S_{ij}(k) c_i(k) \e^{\ri k \tx_{12}} + \S^\ast_{ij}(k) c_j(k) \e^{-\ri k \tx_{12}} + 
\sum_{l=1}^n \S^\ast_{il} (k) c_l(k) \S_{lj}(k) \e^{-\ri k x_{12}} \right ] , 
\label{corr12}
\end{eqnarray}
where $t_{12} = t_1-t_2$, $x_{12} = x_1-x_2$ and $\tx_{12} = x_1+x_2$. 
The invariance of (\ref{corr11}, \ref{corr12}) under time translation implies energy conservation. For 
systems away from equilibrium one expects that the time reversal (\ref{tr1}) symmetry is instead broken. In fact, 
\begin{equation} 
\langle \psi^*(t_1,x_1,i) \psi(t_2,x_2,j)\rangle_{\beta, \mu} \not= \\
\langle \psi^*(-t_2,x_2,j) \psi(-t_1,x_1,i)\rangle_{\beta, \mu} \, .
\label{T2}
\end{equation} 
showing that $T \Omega_{\beta,\mu} \not= \Omega_{\beta,\mu}$, i.e. that $\Omega_{\beta,\mu}$ breaks down 
spontaneously time reversal invariance even if (\ref{tr2}) holds. 
\bigskip

\subsection{Charge transport and density}
\medskip 

Using the explicit form of the correlation function (\ref{corr11}) and applying 
standard point-splitting technique, one obtains 
\begin{eqnarray}
J_i(\beta, \mu ) \equiv \langle j_x(t,x,i) \rangle_{\beta, \mu} = 
\qquad \qquad \qquad \qquad \qquad \qquad \nonumber \\
\lim_{ t_1 \to t_2=t \atop{x_1 \to x_2=x}} \ri \frac{e}{2m} 
\left [ \langle \psi^*(t_1,x_1,i) \prt_{x_2}\psi(t_2,x_2,i)\rangle_{\beta, \mu} - 
\langle \prt_{x_1}\psi^*(t_1,x_1,i) \psi(t_2,x_2,i)\rangle_{\beta, \mu} \right ] = \nonumber \\
 \frac{e}{m} \int_0^\infty \frac{\rd k}{2\pi} k \sum_{j=1}^n \left [\delta_{ij} -  
|\S_{ij}(k)|^2\right ] d_j(k)\, , 
\qquad \qquad \qquad \qquad 
\label{f1}
\end{eqnarray} 
where $\S(k)$ is given by (\ref{S1}) and covers all point-like interactions leading to a self-adjoint 
Schr\"odinger Hamiltonian on the star graph. The current $J_i(\beta, \mu)$ is 
$t$-independent, implying that $\Omega_{\beta, \mu}$ are indeed steady states. We emphasize that (\ref{f1}) 
is precisely the B\"uttiker multi-channel generalization \cite{bu-86} of the 
Landauer expression \cite{la-57} for the steady current. It is also worth stressing that in our context (\ref{f1}) 
is an exact formula, which does not relay on linear response theory. These remarkable features of the states 
$\Omega_{\beta, \mu}$ confirm their physical relevance and suggest to call them L-B states. 

Let us summarize some of the basic properties of the steady current $J_i(\beta, \mu )$. First of all $J_i(\beta, \mu )$ 
is homogeneous ($x$-independent) and,  because of 
unitarity (\ref{unit}), satisfies    
\begin{equation}
\sum_{i=1}^n J_i(\beta, \mu ) = 0 \, .  
\label{k1}
\end{equation}
This is the manifestation of the operator Kirchhoff rule (\ref{KK1}) 
and represents a non-trivial check on (\ref{f1}). Moreover, 
there are two particular cases in which the system is in equilibrium and the current (\ref{f1}) 
must therefore vanish. The first one is when all thermal reservoirs are equivalent 
($\beta_1 = \beta_2 = \cdots = \beta_n$ and 
$\mu_1 = \mu_2 = \cdots = \mu_n$). In fact, (\ref{unit}) implies in this case 
\begin{equation}
J_i(\beta, \mu ) = 0  \, . 
\label{eq11}
\end{equation} 
Another possibility to be in equilibrium is when all transmission coefficients vanish and the leads 
are therefore isolated. In this case $\S(k)$ is diagonal, 
\begin{equation} 
\S_{ij}(k) = \delta_{ij} \e^{\ri \phi_j(k)}\, , \quad \phi_j(k) \in \RR 
\label{diag1}
\end{equation}
which implies (\ref{eq11}) as well. 

In order to illustrate the role of the $\S$-matrices (\ref{S1}), 
it is instructive to consider (\ref{f1}) for $n=2$. Using the general expression (\ref{NN0}), one finds 
\begin{equation}
J_1(\beta, \mu) = -J_2(\beta, \mu) = 
\frac{e}{m}[(\eta_1-\eta_2)\sin(\theta)]^2 \int_0^\infty \frac{\rd k}{2\pi} \frac{k^3}{(k^2+\eta_1^2)(k^2+\eta_2^2)}[d_1(k)-d_2(k)]\, , 
\label{ill1}
\end{equation}
where the sign-difference between $J_1$ and $J_2$ reflects the orientation of the leads. The $k$-integration 
in (\ref{ill1}) can not be performed in a closed analytic form, but the integral is well-defined and can be computed numerically. 
The contour plots of $J_1$ for fixed $e, \theta$ and $m$, displayed in Fig. 2, give an idea about the 
behavior of (\ref{ill1}) in the variables $(\beta_1,\beta_2)$, $(\mu_1,\mu_2)$ and $(\eta_1,\eta_2)$. 
As usual, higher regions are shown in lighter shades. 
The plot on the left concerns $J_1$ in the plane $(\beta_1,\beta_2)$ for fixed $(\mu_1,\mu_2)$ and $(\eta_1,\eta_2)$. 
The plot in the middle illustrates the behavior of $J_1$ as a function of 
$(\mu_1,\mu_2)$, the variables $(\beta_1,\beta_2)$ and $(\eta_1,\eta_2)$ 
being fixed. Finally, the plot on the right shows the dependence on the 
$\S$-matrix variables $(\eta_1,\eta_2)$ at fixed temperatures and chemical potentials. 
\begin{figure}[ht]
\begin{center}
\includegraphics[scale=0.63]{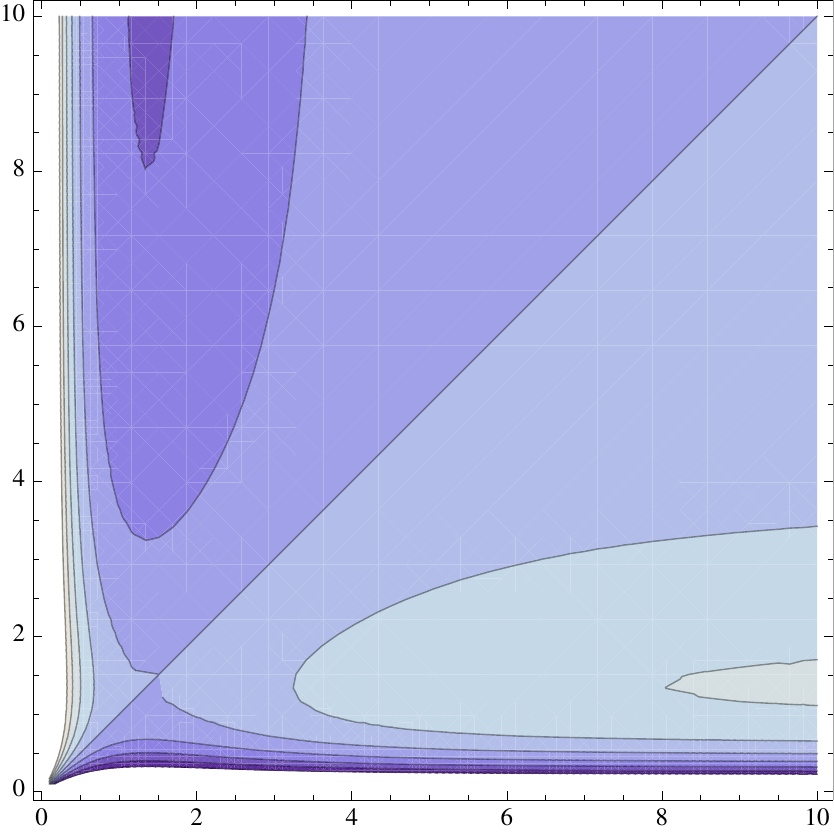}
\hskip 0.11 truecm
\includegraphics[scale=0.63]{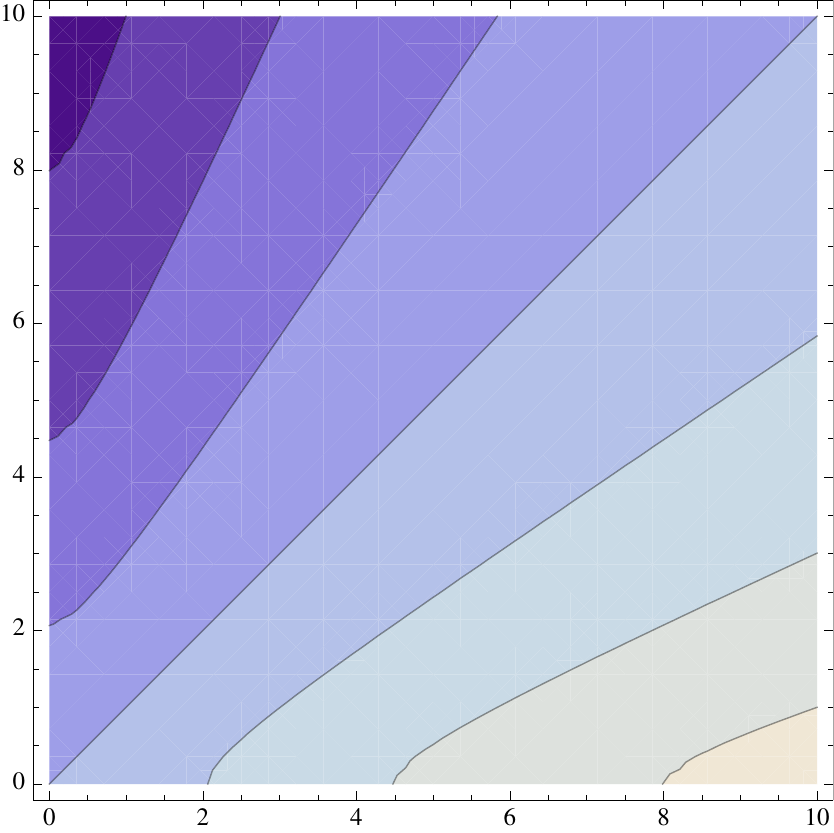}
\hskip 0.11 truecm
\includegraphics[scale=0.63]{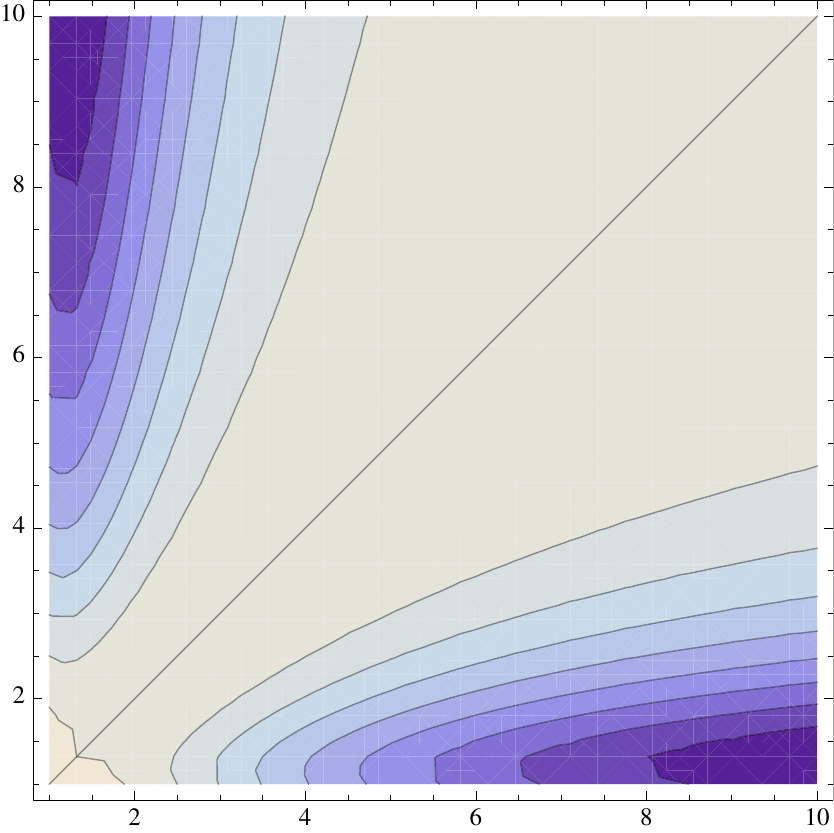}
\end{center}
\centerline{Figure 2: Contour plots of the current $J_1$ in the plane $(\beta_1,\beta_2)$, $(\mu_1,\mu_2)$ and $(\eta_1,\eta_2)$ respectively.} 
\end{figure} 

The plots in Fig. 3 are obtained from those in Fig. 2 by fixing $\beta_1$, $\mu_1$ and $\eta_1$ respectively. 
The sign change of $J_1$ in Fig. 3 indicates that varying $\beta_2$, $\mu_2$ and $\eta_2$ 
one can invert the direction of the current.  
\begin{figure}[ht]
\begin{center}
\includegraphics[scale=0.65]{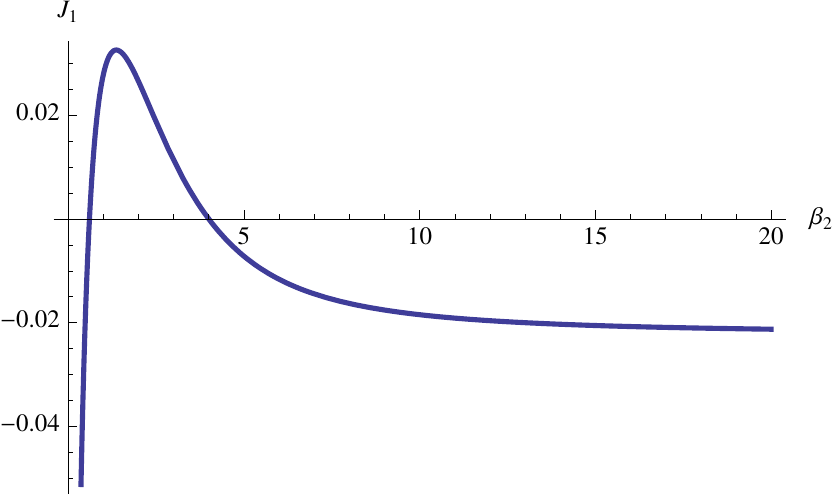}
\hskip 0.1 truecm
\includegraphics[scale=0.65]{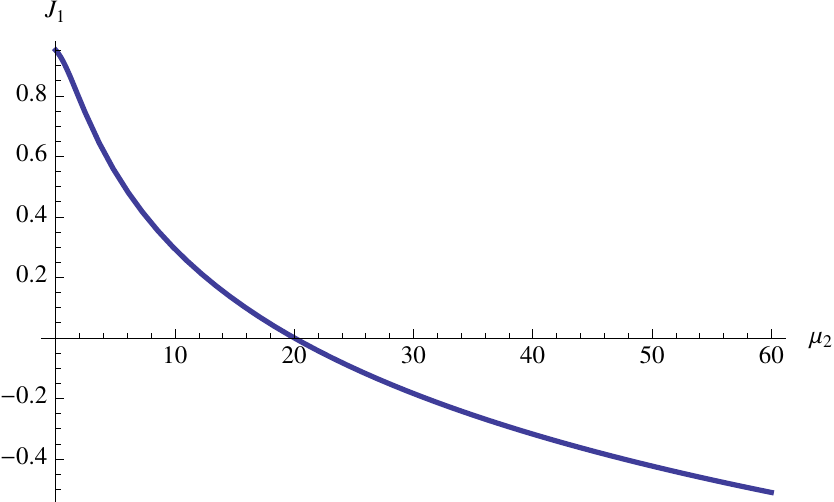}
\hskip 0.1 truecm
\includegraphics[scale=0.65]{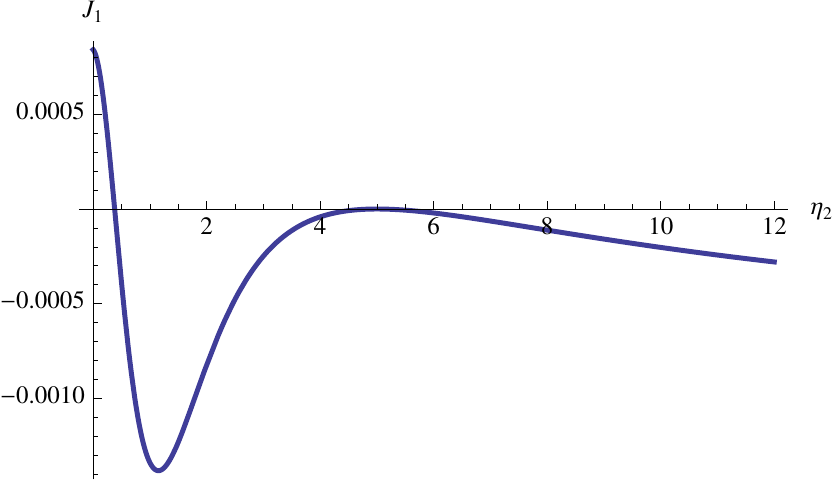}
\end{center}
\centerline{Figure 3: Plots of $J_1$ as a function of $\beta_2$, $\mu_2$ and $\eta_2$ respectively, 
with all other variables fixed.}
\end{figure} 

The expression (\ref{f1}) significantly simplifies at criticality, 
i.e. for {\it scale invariant} point-like interactions in the vertex of $\Gamma$. 
In this case the scattering matrix takes the form 
\begin{equation} 
\S(k) = \theta (k) \UU + \theta(-k) \UU^{-1} \, ,  
\label{crit1} 
\end{equation} 
$\theta$ being the Heaviside step function. Plugging (\ref{crit1}) in (\ref{f1}), one gets 
\begin{equation}
J_i(\beta, \mu ) = 
\frac{e}{2\pi} \sum_{j=1}^n \left (\delta_{ij} - |\UU_{ij}|^2 \right ) \frac{1}{\beta_j} \ln \left (1+ \e^{\beta_j \mu_j} \right ) \, .
\label{f2} 
\end{equation} 
The high and zero temperature limits of (\ref{f2}) are 
\begin{equation}
J_i(0, \mu ) \equiv  \lim_{\beta_k=\beta \to 0} J_i(\beta, \mu ) = 
\frac{e}{4\pi } \sum_{j=1}^n \left (\delta_{ij} - |\UU_{ij}|^2 \right ) \mu_j \, ,  
\label{f3h}
\end{equation} 
\begin{equation}
J_i(\infty, \mu ) \equiv  \lim_{\beta_k=\beta \to \infty} J_i(\beta, \mu ) = 
\frac{e}{2\pi} \sum_{j=1}^n \left (\delta_{ij} - |\UU_{ij}|^2 \right ) \mu_j \, ,  
\label{f4}
\end{equation} 
respectively and are related by 
\begin{equation} 
J_i(0, \mu ) = \frac{1}{2}\, J_i(\infty, \mu ) \, . 
\label{rel}
\end{equation}

By means of (\ref{f1}) one can derive the conductance tensor $\G_{ij}(\beta, \mu )$ defined by 
\begin{equation}
J_i(\beta, \mu ) = \sum_{j=1}^n \G_{ij}(\beta, \mu ) V_j\, , 
\label{cond11}
\end{equation}
where $V_j$ is the voltage applied at the edge $E_j$. It is well known \cite{il-99} that $\G_{ij}(\beta, \mu )$ depends on the 
point on $E_j$ where the voltage is applied. Assuming that this point is deeply in the reservoir with chemical potential $\mu_j$, 
one has \cite{il-99} 
\begin{equation}
V_j = \frac{\mu_j}{e} \, .  
\label{volt}
\end{equation} 
Combining (\ref{f1}), (\ref{cond11}) and (\ref{volt}) one obtains 
\begin{equation} 
\G_{ij}(\beta, \mu ) =  \frac{e^2}{m} \int_0^{\infty}\frac{\rd k}{2\pi} \frac{k}{\mu_j} \left [\delta_{ij} -  
|\S_{ij}(k)|^2\right ] d_j(k)\, ,  
\label{cond12}
\end{equation}
which satisfies Kirchhoff's rule 
\begin{equation}
\sum_{i=1}^n \G_{ij}(\beta, \mu ) = 0  
\label{k2}
\end{equation} 
as it should be. 

Let us focus now on the charge density distribution $\langle j_t(t,x,i)\rangle_{\beta, \mu}$ 
in the state $\Omega_{\beta,\mu}$, which can be computed following the above procedure as well. 
As expected from the current conservation (\ref{cons1}), the charge density 
\begin{eqnarray}
\varrho_i (\beta, \mu, x) \equiv \langle j_t(t,x,i) \rangle_{\beta, \mu} = 
\qquad \qquad \qquad \nonumber \\
e \int_0^{\infty} \frac{\rd k}{2\pi} \left \{ \left [ \S_{ii}(k) \e^{-2\ri k x} +
{\overline \S}_{ii}(k) \e^{2\ri k x} + 1\right ] d_i(k) +
\sum_{j=1}^n |\S_{ij}(k)|^2  d_j(k) \right \}\, , 
\label{f6}
\end{eqnarray} 
is time independent as well. There are however two essential novelties with respect to the current: 
\bigskip

(a) $\varrho_i$ does not vanish at equilibrium; 
\medskip 

(b) $\varrho_i$ depends on the position $x$. 
\bigskip 

\noindent Concerning point (a), we observe that at equilibrium ($n$ isolated leads  with $\S(k)$ defined by (\ref{diag1})) 
one has 
\begin{equation}
\varrho_i^{\rm eq} (\beta, \mu, x) = 
e \int_0^{\infty} \frac{\rd k}{2\pi} \left \{ \left [ \S_{ii}(k) \e^{-2\ri k x} +
{\overline \S}_{ii}(k) \e^{2\ri k x} + 2\right ] d_i(k)\right \}\, . 
\label{f61}
\end{equation} 
Therefore, the non-equilibrium charge distribution is $x$-independent and is given by 
\begin{equation} 
\varrho_i^{\rm neq} (\beta, \mu) \equiv \varrho_i^{\rm eq} (\beta, \mu, x) - \varrho_i (\beta, \mu, x) = 
e \int_0^{\infty} \frac{\rd k}{2\pi} \sum_{j=1}^n\left [\delta_{ij} - |\S_{ij}(k)|^2\right ]  d_j(k)\, . 
\label{f62}
\end{equation} 

The $x$-dependence, mentioned in point (b), is carried by 
\begin{equation}
\varrho_i^{\rm osc} (\beta, \mu ; x) = 
e \int_0^{\infty} \frac{\rd k}{2\pi} \left [ \S_{ii}(k) \e^{-2\ri k x} +
{\overline \S}_{ii}(k) \e^{2\ri k x}\right ] d_i(k)  
\label{FO2}
\end{equation} 
which oscillates with the distance $x$ from the vertex. The appearance of such Friedel-type 
oscillations \cite{fri-52} confirms once more that the junction behaves 
indeed as a point-like defect. Since the integration in (\ref{FO2}) can not be performed in closed form, 
in order to get an idea about the oscillations, it is useful to consider the zero-temperature limit 
\begin{equation} 
\varrho_i^{\rm osc} (\infty, \mu ; x) \equiv 
\lim_{\beta_k=\beta \to \infty} 
\varrho_i^{\rm osc} (\beta, \mu ; x) = 
e \int_0^{\sqrt {2m\mu_i}} \frac{\rd k}{2\pi} \left [ \S_{ii}(k) \e^{-2\ri k x} +
{\overline \S}_{ii}(k) \e^{2\ri k x}\right ] \, . 
\label{FO3}
\end{equation} 
At criticality (\ref{crit1}) and setting $\UU_{ii} = \overline{\UU}_{ii}$ for simplicity, one finds 
\begin{equation}
\varrho_i^{\rm osc} (\infty, \mu ; x) = 
\frac{e\UU_{ii}}{\pi x} \sin(2x\sqrt {2m\mu_i}) \, ,
\label{FO4}
\end{equation} 
which shows that the amplitude of the oscillations on the graph decays with the distance 
from the vertex like $x^{-1}$, which is a typical behavior in one space dimension. 
\bigskip 

\subsection{Casimir effect away from equilibrium and heat flow} 
\medskip 

The structure of the energy density in the state $\Omega_{\beta , \mu}$ resembles very much that of the 
charge density (\ref{f6}). One has   
\begin{eqnarray}
{\cal E}_i(x; \beta, \mu ) \equiv \langle \theta_{tt}(t,x,i) \rangle_{\beta, \mu} = 
\qquad \qquad \qquad \qquad \qquad \nonumber \\
\frac{1}{2}
\int_0^{\infty} \frac{\rd k}{2\pi} \omega (k) \left \{ \left [ \S_{ii}(k) \e^{-2\ri k x} +
{\overline \S}_{ii}(k) \e^{2\ri k x} + 1\right ] d_i(k) +
\sum_{j=1}^n |\S_{ij}(k)|^2  d_j(k) \right \}\, , 
\label{f7}
\end{eqnarray} 
This result confirms the presence of Friedel oscillations in the energy density as well. 
It is instructive to compare (\ref{f7}) to the equilibrium energy density  
\begin{equation}
{\cal E}_i^{\rm eq} (x; \beta, \mu ) = 
\int_0^{\infty} \frac{\rd k}{2\pi} \omega (k) \left [ \S_{ii}(k) \e^{-2\ri k x} +
{\overline \S}_{ii}(k) \e^{2\ri k x} + 2\right ] d_i(k) \, ,  
\label{f7e}
\end{equation} 
associated to (\ref{diag1})). One finds  
\begin{equation}
{\cal E}_i^{\rm eq} (x; \beta, \mu ) - {\cal E}_i(x; \beta, \mu )= \frac{1}{2}
\int_0^{\infty} \frac{\rd k}{2\pi} \omega (k) \sum_{j=1}^n\left [\delta_{ij} - |\S_{ij}(k)|^2\right ]  d_j(k) \, , 
\label{f7ne}
\end{equation} 
which gives the genuine non-equilibrium part of the energy density (\ref{f7}). 
The $x$-independent contribution to (\ref{f7}), namely  
\begin{equation}
\varepsilon_i(\beta, \mu ) = 
 \frac{1}{2}\sum_{j=1}^n \int_0^{\infty} \frac{\rd k}{2\pi} \omega (k) \left [ \delta_{ij} +  |\S_{ij}(k)|^2 \right ] d_j(k) 
\label{SB}
\end{equation} 
is the Stefan-Boltzmann law in the present context. At criticality 
\begin{equation}
\varepsilon_i(\beta, \mu ) = -\frac{1}{4} \sqrt{\frac{m}{2\pi}} 
\sum_{j=1}^n \left ( \delta_{ij} + |\UU_{ij}|^2  \right ) 
\frac{1}{\beta_j^{\frac{3}{2}}} {\rm Li}_{\frac{3}{2}} \left (-\e^{\beta_j \mu_j} \right )\, , 
\label{critSB}
\end{equation} 
where ${\rm Li}_s$ is the polylogarithm function.

The counterpart of the L-B formula for the heat (energy) flow is  
\begin{equation}
\langle \theta_{xt}(t,x,i) \rangle_{\beta, \mu} = 
\frac{1}{m} \int_0^{\infty} \frac{\rd k}{2\pi} k\, \omega (k) \sum_{j=1}^n \left [\delta_{ij} -  
|\S_{ij}(k)|^2\right ] d_j(k) \equiv 
{\cal T}_i(\beta, \mu )  \, . 
\label{f8}
\end{equation} 
Apart from the additional $\omega (k)$ factor in the integrand of (\ref{f8}), the charge (\ref{f1}) and energy (\ref{f8}) flows 
have the same structure. For this reason ${\cal T}_i(\beta , \mu)$ shares with $J_i(\beta , \mu)$  
the general properties listed after equation (\ref{f1}). In the scale invariant case the energy flow is 
\begin{eqnarray}
{\cal T}_i(\beta, \mu ) = 
\frac{1}{2m^2} \sum_{j=1}^n \left (\delta_{ij} - |\UU_{ij}|^2 \right ) 
\int_0^{\infty} \frac{\rd k}{2\pi} \frac{k^3 \e^{-\beta_j [\omega (k) - \mu_j]}}{1 + \e^{-\beta_j [\omega (k) - \mu_j]}} = 
\nonumber \\
\frac{1}{2\pi} \sum_{j=1}^n \left (|\UU_{ij}|^2 - \delta_{ij} \right ) \frac{1}{\beta_j^2} {\rm Li}_2(-\e^{\beta_j \mu_j}) \, , 
\qquad \qquad \qquad 
\label{f9}
\end{eqnarray} 
which gives in the zero temperature limit 
\begin{equation}
{\cal T}_i(\infty,\mu ) \equiv \lim_{\beta_k=\beta \to \infty}=  \frac{1}{4\pi} \sum_{j=1}^n \left (\delta_{ij} - |\UU_{ij}|^2 \right ) \mu_j^2\, .  
\label{f10}
\end{equation} 
\bigskip 

\subsection{Noise} 
\medskip 

In this section we derive the {\it noise power} generated by the point like interactions in the Schr\"odinger 
junction. For this purpose we need \cite{mla-92}-\cite{bb-00} 
the two-point {\it connected} current-current correlator. 
After some algebra one finds  
\begin{eqnarray}
\langle j_x(t_1,x_1,i) j_x(t_2,x_2,j) \rangle_{\beta, \mu}^{\rm conn} \equiv 
\langle j_x(t_1,x_1,i) j_x(t_2,x_2,j) \rangle_{\beta, \mu} - 
\langle j_x(t_1,x_1,i)\rangle_{\beta, \mu}\langle j_x(t_2,x_2,j) \rangle_{\beta, \mu} = \nonumber \\
-\frac{e^2}{4m^2} \int_{-\infty}^0 \frac{\rd k_1}{2\pi}  \int_{-\infty}^0 \frac{\rd k_2}{2\pi} \e^{\ri t_{12} [\omega(k_1) - \omega(k_2)]} 
\sum_{l,m=1}^n d_l (k_1) c_m (k_2) \qquad \qquad \qquad \qquad \quad 
\nonumber \\
\times \Bigl \{ \chi^*_{li}(k_1;x_1) \left [\der_{x} \chi_{i m}\right ](k_2;x_1) - 
\left [\der_{x} \chi^*_{l i}\right ](k_1;x_1) \chi_{im}(k_2;x_1) \Bigr \} \qquad \qquad \qquad \qquad \, \nonumber \\
\times \Bigl \{ \chi^*_{mj}(k_2;x_2) \left [\der_{x} \chi_{j l}\right ](k_1;x_2) - 
\left [\der_{x} \chi^*_{m j}\right ](k_2;x_2) \chi_{jl}(k_1;x_2) \Bigr \}\, , \qquad \qquad \qquad \quad 
\label{N1}
\end{eqnarray}
where $\chi$ and $\chi^\ast$ are given by (\ref{KK3}). Using the time translation 
invariance of (\ref{N1}), the noise power is defined \cite{bb-00} by 
\begin{equation}
P_{ij}(\beta, \mu ; x_1, x_2 ; \omega) \equiv \int_{-\infty}^\infty \rd t\,  \e^{i \omega t} \, 
\langle j_x(t,x_1,i) j_x(0,x_2,j) \rangle_{\beta, \mu}^{\rm conn}
\label{N2}
\end{equation}
The zero-frequency limit ({\it zero-frequency noise power}) 
\begin{equation} 
P_{ij}(\beta, \mu ) \equiv \lim_{\omega \to 0^+} P_{ij}(\beta, \mu ; x_1, x_2 ; \omega) 
\end{equation}
turns out to be $x_{1,2}$-independent and is given by: 
\begin{eqnarray}
P_{ij}(\beta, \mu ) = \frac{e^2}{m} \int^{\infty}_0 \frac{\rd k}{2 \pi} k 
\Biggl [ \delta_{ij} d_{i}(k) c_{i}(k) - 
|\S_{ij}(k)|^2 d_{j}(k) c_{j}(k) - |\S_{ji}(k)|^2 d_{i}(k) c_{i}(k) + \nonumber \\
+\sum_{l,m=1}^n \S_{il}(k)c_l(k)\overline{\S}_{jl}(k) \S_{jm}(k)d_m(k)\overline{\S}_{im}(k) \Biggr ]\, . 
\qquad \qquad \qquad \qquad 
\label{N3}
\end{eqnarray}  
It is instructive to summarize at this point the general features of (\ref{N3}): 
\medskip 

(i) $P_{ij}(\beta, \mu )$ is symmetric in $i$ and $j$. The first three terms of the integrand are manifestly symmetric. 
Concerning the last term, using the identity $c_i (k) = 1- d_i(k)$ one gets 
\begin{eqnarray}
\sum_{l,m=1}^n \S_{il}(k)c_l(k)\overline{\S}_{jl}(k) \S_{jm}(k)d_m(k)\overline{\S}_{im}(k) = 
\qquad \qquad \qquad \nonumber \\
\delta_{ij} \sum_{m=1}^n\S_{im}(k)d_m(k)\overline{\S}_{im}(k)  \mp  
\sum_{l,m=1}^n \S_{il}(k)d_l(k)\overline{\S}_{jl}(k) \S_{jm}(k)d_m(k)\overline{\S}_{im}(k) \, , 
\label{N4}
\end{eqnarray}
which is symmetric as well. One can therefore rewrite $P_{ij}(\beta, \mu )$ in the following manifestly symmetric form: 
\begin{eqnarray}
P_{ij}(\beta, \mu ) = \frac{e^2}{m} \int^{\infty}_0 \frac{\rd k}{2 \pi} k 
\Bigl \{ \delta_{ij} d_{i}(k) c_{i}(k) - 
|\S_{ij}(k)|^2 d_{j}(k) c_{j}(k) - |\S_{ji}(k)|^2 d_{i}(k) c_{i}(k) + 
\nonumber \\
+\frac{1}{2}\sum_{l,m=1}^n \S_{il}(k)\overline{\S}_{jl}(k) 
\S_{jm}(k)\overline{\S}_{im}(k) [c_l(k)d_m(k)+c_m(k)d_l(k)]\Bigr \}\, . 
\qquad \qquad \qquad 
\label{N5}
\end{eqnarray}

\medskip

(ii) The last identity implies also that $P_{ij}(\beta, \mu )$ is real; 
\medskip 

(iii) As expected, $P_{ij}(\beta, \mu )$ satisfies the Kirchhoff rule
\begin{equation}
\sum_{i=1}^n P_{ij}(\beta, \mu ) = \sum_{j=1}^n P_{ij}(\beta, \mu ) = 0 \, ,  
\label{N6}
\end{equation}
which provides an useful check. One has actually 
\begin{equation}
\sum_{i=1}^n P_{ij}(\beta, \mu ; 0, x_2 ; \omega)= \sum_{j=1}^n P_{ij}(\beta, \mu ; x_1, 0 ; \omega) = 0 \, ,  
\label{N7}
\end{equation}
at any frequency $\omega$. 
\medskip 

(iv) All noise components $P_{ij}(\beta, \mu )$ vanish for isolated leads (\ref{diag1}). 
\bigskip 

Let us discuss now the behavior of the noise, starting with the case $n=2$. Combining (\ref{NN0}) with 
(\ref{N5}), we find 
\begin{eqnarray} 
P_{11}(\beta,\mu) = [e(\eta_1-\eta_2)\sin(\theta)]^2 \frac{1}{m} \int_0^\infty \frac{\rd k}{2\pi} 
\frac{k^3}{(k^2+\eta_1^2)(k^2+\eta_2^2)} \qquad \quad 
\nonumber \\
\left \{d_1(k)+d_2(k) - 2d_1(k)d_2(k) -
\frac{k^2[(\eta_1-\eta_2)\sin(\theta)]^2}{(k^2+\eta_1^2)(k^2+\eta_2^2)} [d_1(k)-d_2(k)]^2\right \}\, . 
\label{a1}
\end{eqnarray}

{}For $\eta_1 = \eta_2$ and/or $\theta=0$ the leads are isolated (see (\ref{NN0})) 
and the noise vanishes according to point (iv) above. 
Like for the steady current, we report some contour plots, showing the complicated dependence 
of the noise on the parameters $(\beta_1,\beta_2)$, $(\mu_1,\mu_2)$ and 
$(\eta_1,\eta_2)$ for fixed $e, \theta$ and $m$. Fig. 4 illustrates the 
behavior of $P_{11}(\beta,\mu) $ in each pair of these variables, 
the remaining two being fixed. The left plot is the noise in the plane $(\beta_1,\beta_2)$. 
In the middle we display the noise as a function of the chemical potentials $(\mu_1,\mu_2)$. 
Finally, the right plot shows the dependence on the $\S$-matrix variables $(\eta_1,\eta_2)$.  

\begin{figure}[ht]
\begin{center}
\includegraphics[scale=0.63]{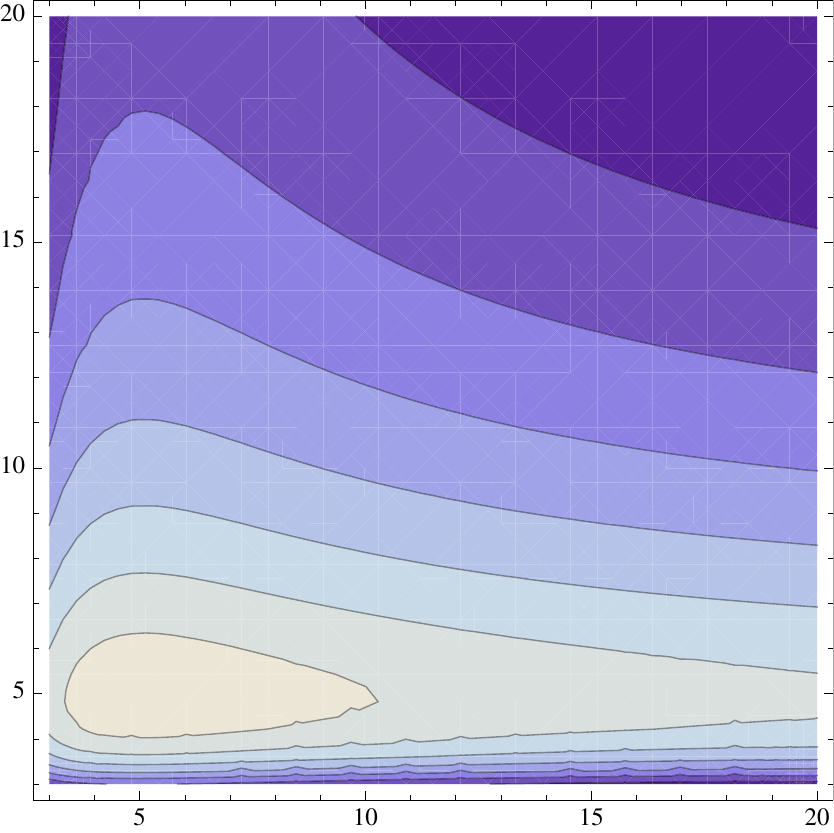}
\hskip 0.11 truecm
\includegraphics[scale=0.63]{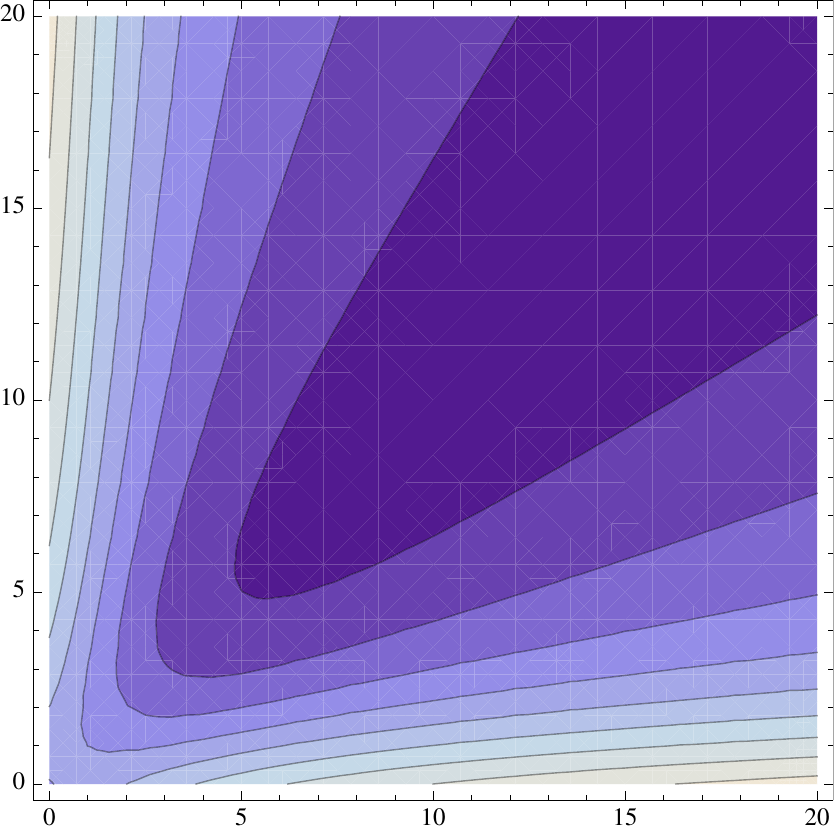}
\hskip 0.11 truecm
\includegraphics[scale=0.63]{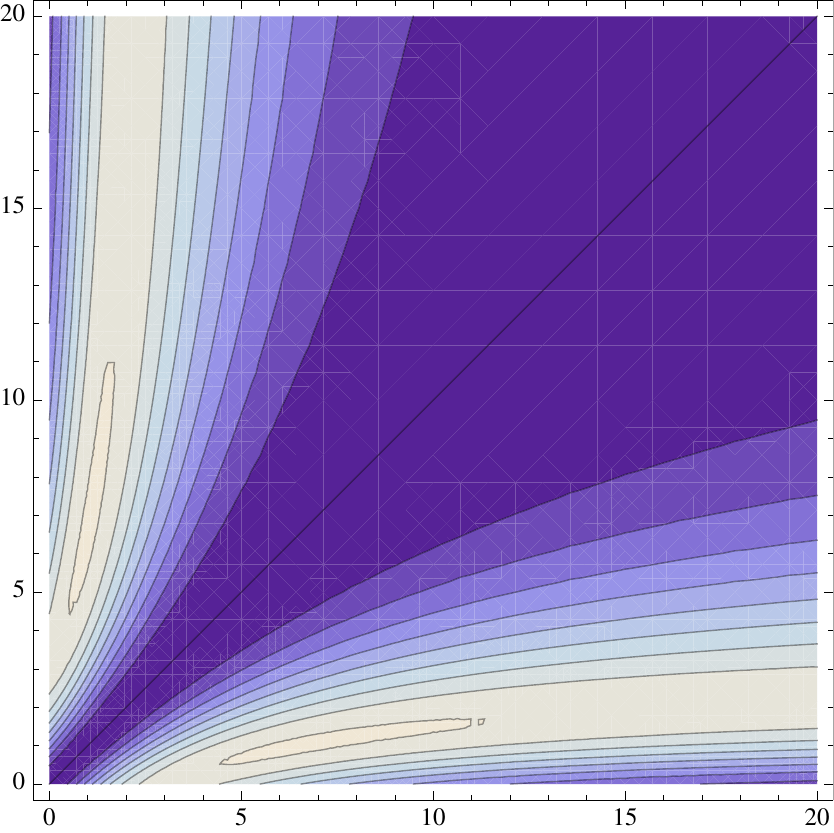}
\end{center}
\centerline{Figure 4: Contour plots of the noise $P_{11}$ in the plane $(\beta_1,\beta_2)$, $(\mu_1,\mu_2)$ and $(\eta_1,\eta_2)$ respectively.} 
\end{figure} 

\begin{figure}[ht]
\begin{center}
\includegraphics[scale=0.65]{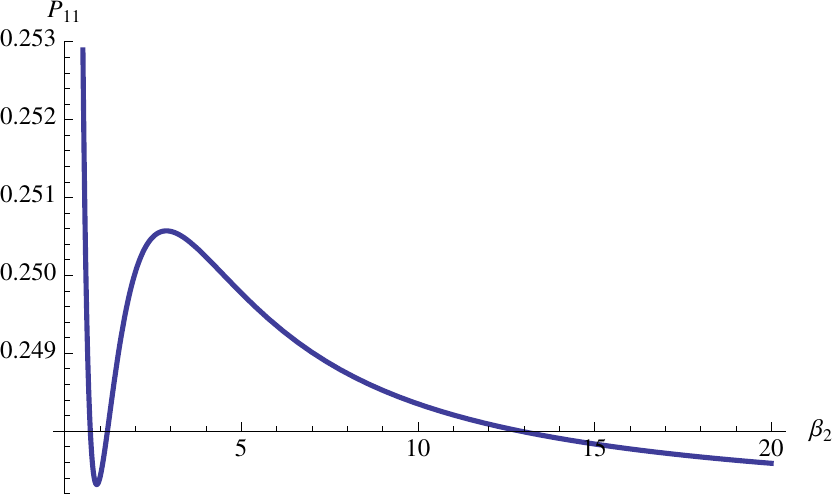}
\hskip 0.1 truecm
\includegraphics[scale=0.65]{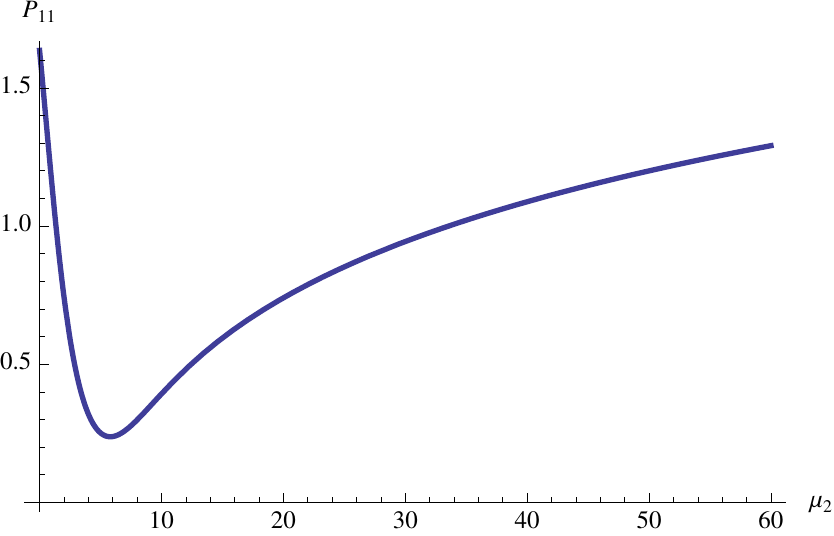}
\hskip 0.1 truecm
\includegraphics[scale=0.65]{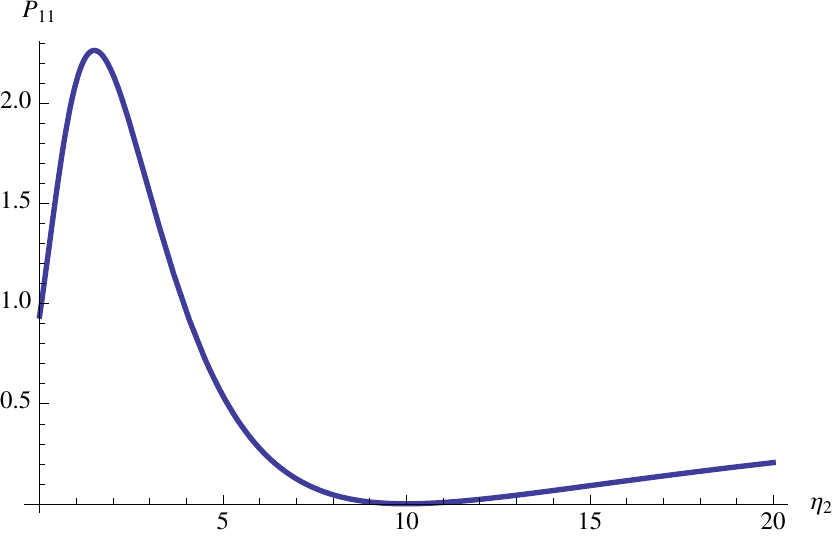}
\end{center}
\centerline{Figure 5: Plots of $P_{11}$ as a function of $\beta_2$, $\mu_2$ and $\eta_2$ respectively, 
with all other variables fixed.} 
\end{figure} 

The plots in Fig. 5, obtained from those in Fig. 4 by fixing $\beta_1$, $\mu_1$ and $\eta_1$ respectively, 
confirm that the noise $P_{ij}(\beta, \mu)$ depends in a complicated way on 
$\beta$, $\mu$ and the $\S$-matrix parameters. At criticality however, as expected on general grounds, 
the situation simplifies and one can push further the analytic computation. In fact, inserting 
(\ref{crit1}) in (\ref{N5}) one gets, 
\begin{eqnarray} 
P_{ij}(\beta, \mu ) = \frac{e^2}{m} 
\Biggl \{ \delta_{ij} I_{ii} (\beta) - 
|\UU_{ij}|^2 I_{jj} (\beta) - |\UU_{ji}|^2 I_{ii} (\beta) + 
\nonumber \\
+\frac{1}{2}\sum_{l,m=1}^n \overline{\UU}_{il}\, \UU_{jl}\, \overline{\UU}_{jm}\, \UU_{im}\, 
[I_{lm} (\beta) + I_{ml} (\beta)] \Biggr \}\, , \qquad \quad 
\label{N11}
\end{eqnarray}
where 
\begin{equation}
I_{ij} (\beta) \equiv \int^{\infty}_0 \frac{\rd k}{2 \pi}\, k\, d_i(k) c_j (k)\, . 
\label{a2}
\end{equation} 
For equal temperatures $\beta_i=\beta_j=\beta$ the integration in (\ref{a2}) can be performed 
explicitly and one finds 
\begin{equation}
I_{ij} (\beta)  = 
\begin{cases} 
\frac{m}{2\pi \beta} \frac{\e^{\beta \mu}}{1+\e^{\beta \mu}}   & \qquad  \text{if\; $\mu_i=\mu_j\equiv \mu$}\, , \\
\frac{m}{2\pi \beta} \frac{\e^{\beta \mu_i}}{\e^{\beta \mu_j} - \e^{\beta \mu_i}}
\ln \left (\frac{1+\e^{\beta \mu_j}}{1+\e^{\beta \mu_i}}\right ) & \qquad \text{if\; $\mu_i\not=\mu_j$}\, . \\
\end{cases}
\label{N12}
\end{equation} 
Therefore, in the case $\beta_1=\beta_2=\cdots =\beta_n = \beta$ with generic chemical potentials $\mu_i \geq 0$ one obtains 
\begin{eqnarray}
P_{ij}(\beta, \mu ) = \frac{e^2}{2\pi \beta}
\Biggl [ \delta_{ij} \frac{\e^{\beta \mu_{i}}}{1+\e^{\beta \mu_{i}}} - 
|\UU_{ij}|^2  \frac{\e^{\beta \mu_{j}}}{1+\e^{\beta \mu_{j}}} - |\UU_{ji}|^2  \frac{\e^{\beta \mu_{i}}}{1+\e^{\beta \mu_{i}}} + 
\qquad \quad \nonumber \\
+\sum_{l=1}^n |\UU_{il}|^2 |\UU_{jl}|^2 \frac{\e^{\beta \mu_l}}{1+\e^{\beta \mu_l}}
+\frac{1}{2}\sum_{l,m=1\atop l\not=m}^n \overline{\UU}_{il} \UU_{jl} \overline{\UU}_{jm}\UU_{im} 
\frac{\e^{\beta \mu_l} +\e^{\beta \mu_m}}{\e^{\beta \mu_l} - \e^{\beta \mu_m}}
\ln \left (\frac{1+\e^{\beta \mu_l}}{1+\e^{\beta \mu_m}}\right ) \Biggr ]\, . 
\label{N13}
\end{eqnarray} 

It is instructive to consider at this stage the two limits leading to the shot and thermal noise. For deriving 
the shot noise, we need the $\beta \to \infty$ limit of the integrals (\ref{N12}), which are 
\begin{equation}
\lim_{\beta \to \infty} I_{ij}(\beta) = 
\begin{cases} 
0   & \qquad  \text{if\; $\mu_i \leq \mu_j $}\, , \\
\frac{m}{2\pi} (\mu_i - \mu_j)  & \qquad \text{if\; $\mu_i > \mu_j$}\, . \\
\end{cases}
\label{N14}
\end{equation} 
Therefore, in the scale invariant case (\ref{crit1}) the shot noise is 
\begin{equation}
P_{ij}(\mu )\equiv \lim_{\beta \to \infty} P_{ij}(\beta, \mu ) = \frac{e^2}{4\pi} 
\sum_{l,m=1\atop l\not=m}^n \overline{\UU}_{il} \UU_{jl} \overline{\UU}_{jm}\UU_{im} |\mu_l -\mu_m| \, , 
\label{N15}
\end{equation} 
which exhibits the standard behavior \cite{mla-92}-\cite{bb-00} in terms of $|\mu_i-\mu_j|$. 

In order to compute the thermal noise, we consider (\ref{N13}) for $\mu_1=\mu_2=\cdots =\mu_n = \mu$. 
One has 
\begin{equation}
P_{ij}(\beta, \mu ) =\frac{e^2}{2\pi \beta}  \frac {\e^{\beta \mu}}
{1+ \e^{\beta \mu}} \left [2\delta_{i j} - |\UU_{ij}|^2 - |\UU_{ji}|^2\right ] \, .
\label{N9}
\end{equation}
In the purely thermal case ($\mu  \to 0$) one finds  
\begin{equation}
P_{ij}(\beta) =\frac{e^2}{2\pi \beta}  \left [2\delta_{i j} - |\UU_{ij}|^2 - |\UU_{ji}|^2\right ] 
\sim T \, ,  
\label{N10}
\end{equation}
which is the well-known Johnson-Nyquist formula. 

A remarkable feature of the thermal noise is that away from criticality 
the point-like interactions at the vertex can modify the linear behavior (\ref{N10}) for large $T$. 
Let us consider indeed (\ref{a1}) for $\beta_1 = \beta_2 = \beta$ and $\mu_1=\mu_2=0$, namely
\begin{equation}
P_{11}(\beta)  
=2 \left [e (\eta_1-\eta_2) \sin(\theta)\right ]^2 \frac{1}{m} \int_0^\infty \frac{\rd k}{2\pi}
\frac{k^3\e^{-\beta\frac{k^2}{2m}}}
{(k^2+\eta_1^2)(k^2+\eta_2^2)\left (1+\e^{-\beta\frac{k^2}{2m}}\right )^2}\, .
\label{NN1}
\end{equation}
Assuming for simplicity $\eta_2 =0$ and introducing the variables 
$\eta =\eta_1$ and $\xi = \e^{-\beta\frac{k^2}{2m}}$, one gets 
\begin{equation} 
P_{11}(\beta) =  
[e \eta \sin(\theta)]^2 \frac{1}{2\pi m}
\int_0^1 \rd \xi \frac{1}{(1+\xi)^2 \left [\frac{\beta \eta^2}{2m} - \ln (\xi)\right ]}\, .
\label{NN2}
\end{equation}
Since the $k$-integration in (\ref{NN0}) can not be performed exactly, in order to estimate the temperature 
dependence one can use the inequalities 
\begin{equation} 
[e \eta \sin(\theta)]^2 \frac{1}{4\pi m} I \left (\frac{\beta \eta^2}{2m} \right ) 
\leq P_{11}(\beta) \leq [e \eta \sin(\theta)]^2 \frac{1}{2\pi m} I \left (\frac{\beta \eta^2}{2m} \right )  \, , 
\label{NN3}
\end{equation} 
where \cite{gr-07}  
\begin{equation} 
I(a) = \int_0^1 \rd \xi \frac{1}{\left [a-\ln(\xi)\right ]} = - \e^a\, {\rm Ei} (-a)\, , 
\label{NN4}
\end{equation}
${\rm Ei} $ being the exponential integral function. In this way one finds
\begin{equation}
P_{11}(\beta) \sim
\begin{cases} 
T  & \qquad \text{for\; $T\to 0$}\, , \\
\ln (T) & \qquad \text{for\; $T \to \infty$}\, , \\
\end{cases}
\label{NN5}
\end{equation}
which shows that the $k$-dependence of the $\S$-matrix indeed modifies the 
Johnson-Nyquist behavior at high temperatures. The milder logarithmic divergence 
for large $T$ provides an attractive experimental signature. 
\bigskip

\subsection{External electromagnetic field} 
\medskip 

In the above considerations the interaction was localized in the junction. We extend here the framework 
to the more realistic physical situation of a junction in a three-dimensional ambient space with a 
classical static magnetic field, interacting with the Schr\"odinger 
excitations along the leads. The graph $\Gamma$, modeling the junction, 
is embedded in $\RR^3$, equipped with a Cartesian coordinate system whose origin 
$O$ coincides with the vertex $V$ of $\Gamma$. The direction of each edge $E_i \subset \RR^3$ is determined 
by the unit vector $\ebf^{(i)}$. At any point $P\in \RR^3$ the magnetic field ${\bf B}(P) = {\rm rot} [\abf (P)]$ 
is generated by the potential $\abf (P)$. The minimal coupling 
of the Schr\"odinger field $\varphi$ with $\abf$ gives the following equation of motion 
\begin{equation}
\left [\ri \prt_t -\frac{1}{2m} \left (\ri \prt_x - e A_x(x,i)\right )\left (\ri \prt_x - e A_x(x,i)\right )\right ]\varphi (t,x,i) = 0\, , 
\label{m1}
\end{equation}
$A_x(x,i)$ being the projection 
\begin{equation} 
A_x(x,i) = \ebf^{(i)}\cdot \abf (P) \, , \qquad P\equiv (x,i) \in \Gamma \subset \RR^3\, ,  
\label{m2}
\end{equation} 
of the potential $\abf$ along the edge $E_i$. 
All self-adjoint extensions of the relative Hamiltonian are now parametrized by the boundary conditions \cite{ksch-03}  
\begin{equation} 
\lim_{x\to 0^+}\sum_{j=1}^n \left [\lambda (\II-\UU)_{ij} - (\II+\UU)_{ij}(\ri \prt_x - A_x(x,j))\right ] \varphi (t,x,j) = 0\, . 
\label{a3} 
\end{equation} 
The conserved electric current is 
\begin{equation}
j_x(t,x,i)= \ri \frac{e}{2m} \left [ \varphi^*(\partial_x\varphi ) - (\partial_x\varphi^*)\varphi \right ]  (t,x,i) -\frac{e}{m} A_x(x,i)(\varphi^*\varphi)(t,x,i) \, . 
\label{a4}
\end{equation}
It is easy to show now that the solution of the problem (\ref{m1},\ref{a3}) can be reduced to that described in section 3.1. 
Indeed, let us introduce 
\begin{equation} 
\psi(t,x,i) = \e^{-ie\alpha (x,i)} \varphi (t,x,i)\, , \qquad \alpha (x,i) = \int_x^\infty \rd y A_y(y,i) \, , 
\label{a5}
\end{equation} 
where we assumed that $A_x(x,i)$ are integrable on the half line. 
Notice that $\psi$ and $\varphi$ have the same behavior for $x\to \infty$. Moreover, $\psi$ satisfies (\ref{eqm1}) and in terms of $\psi$ 
the current (\ref{a4}) takes precisely the form (\ref{curr1}). The interaction is totally absorbed in the boundary condition for $\psi$ 
following from (\ref{a3},\ref{a5}). One has 
\begin{equation} 
\lim_{x\to 0^+}\sum_{j=1}^n \left [\lambda (\II-\UU(\abf))_{ij} - \ri (\II+\UU(\abf))_{ij}\prt_x \right ] \psi (t,x,j) = 0\, , 
\label{a6} 
\end{equation} 
where 
\begin{equation}
\UU_{ij}(\abf) = \e^{-\ri e \alpha_i}\, \UU_{ij}\, \e^{\ri e \alpha_j}\, , \qquad \alpha_i = \alpha (0,i) \, . 
\label{a7}
\end{equation} 
Combining (\ref{S1}) and (\ref{a7}) one concludes that the substitution 
\begin{equation} 
\S_{ij}(k) \longmapsto \e^{-\ri e \alpha_i}\, \S_{ij}(k) \, \e^{\ri e \alpha_j}  
\label{a8}
\end{equation} 
extends all the results of this section to the case of a junction minimally coupled to a time-independent 
ambient magnetic field. We stress that the correlation functions of fields localized in 
different edges (see e.g. (\ref{corr11},\ref{corr12},\ref{N1})) are sensitive to the transformation (\ref{a8}). 
In particular, the field $\abf$ has a non-trivial impact on the noise power at frequency $\omega \not=0$. 
We will analyze this issue in more details elsewhere.  

\bigskip 

\subsection{Remarks}
\medskip

Let us discuss first the role of possible bound states of $\S(k)$, which have been excluded 
in the above considerations by assuming (\ref{compl1}). It has been shown in 
previous work (\cite{Mintchev:2004jy}, \cite{Bellazzini:2008cs}, \cite{Bellazzini:2010gs}) 
that the bound states generate new 
quantum degrees of freedom, which have a non-trivial contribution to the correlation function (\ref{corr11}). 
The key point is that this contribution depends on the space-time coordinates only through the combinations 
$t_{12}$ and $\tx_{12}$. According to (\ref{curr1}, \ref{en1}), the charge (\ref{f1}) and energy (\ref{f9}) flows 
are therefore not affected by the presence of bound states. The relative densities however get \cite{Mintchev:2004jy} 
nontrivial bound state contributions. 

One can investigate along the above lines also the Schr\"odinger equation (\ref{eqm1}) with Bose statistics. 
The final results in this case obviously follow from equations (\ref{corr11},\ref{f1},\ref{f6}-\ref{f9},\ref{N5}) by 
substituting the Fermi distribution with the Bose distribution $d_i^-(k)$. Concerning the noise, 
in the scale invariant case one gets for bosons  
\begin{eqnarray}
P^-_{ij}(\beta, \mu ) = \frac{e^2}{2\pi \beta}
\Biggl [ \delta_{ij} \frac{\e^{\beta \mu_{i}}}{1-\e^{\beta \mu_{i}}} - 
|\UU_{ij}|^2  \frac{\e^{\beta \mu_{j}}}{1-\e^{\beta \mu_{j}}} - |\UU_{ji}|^2  \frac{\e^{\beta \mu_{i}}}{1-\e^{\beta \mu_{i}}} + 
\qquad \nonumber \\
+\sum_{l=1}^n |\UU_{il}|^2 |\UU_{jl}|^2 \frac{\e^{\beta \mu_l}}{1-\e^{\beta \mu_l}}
+\frac{1}{2}\sum_{l,m=1\atop l\not=m}^n \overline{\UU}_{il} \UU_{jl} \overline{\UU}_{jm}\UU_{im} 
\frac{\e^{\beta \mu_l} +\e^{\beta \mu_m}}{\e^{\beta \mu_l} - \e^{\beta \mu_m}}
\ln \left (\frac{1-\e^{\beta \mu_m}}{1-\e^{\beta \mu_l}}\right ) \Biggr ]\, , 
\label{r1}
\end{eqnarray} 
which, compared to (\ref{N13}), shows haw the zero frequency noise power depends on the statistics. For 
instance, in the shot noise limit one obtains 
\begin{equation}
P^-_{ij}(\mu )\equiv \lim_{\beta \to \infty} P^-_{ij}(\beta, \mu ) = -\frac{e^2}{4\pi} 
\sum_{l,m=1\atop l\not=m}^n \overline{\UU}_{il} \UU_{jl} \overline{\UU}_{i_jm}\UU_{im} |\mu_l -\mu_m| \, ,  
\label{r2}
\end{equation} 
which has the magnitude of the fermionic shot noise (\ref{N15}) but the opposite sign \cite{bb-00}. 
\bigskip

\section{The Dirac junction} 

\subsection{Preliminaries} 
\medskip

The massless Dirac equation on the star graph $\Gamma$ is 
\begin{equation}
(\gamma_t \prt_t - \gamma_x \prt_x)\psi (t,x,i) = 0 \, , 
\qquad x> 0 \, , 
\label{eqm2}
\end{equation} 
where 
\begin{equation} 
\psi (t,x,i)=\left (\begin{array}{c}\psi_1(t,x,i) \\ \psi_2(t,x,i) \\ \end{array}\right )\, , \quad 
\gamma_t = \left(\begin{array}{cc}0 & 1\\ 1 &  0 \\ \end{array} \right)\, , \quad 
\gamma_x = \left(\begin{array}{cc}0 & 1\\ -1 &  0 \\ \end{array} \right)\, .   
\label{not1}
\end{equation}
We assume that $\psi_\alpha $ satisfy the conventional equal-time anti-commutation relations. 
The boundary conditions which define all self-adjoint extensions of the bulk Hamiltonian 
$\ri \gamma_t \gamma_x \prt_x$ are \cite{bh-03}, \cite{bbm-un} 
\begin{equation} 
\psi_1(t,0,i) = \sum_{j=1}^n \UU_{ij} \psi_2(t,0,j) \, , 
\label{bc2}
\end{equation}
where $\UU$ is any unitary $n\times n$ matrix. In physical terms $\UU$ parametrizes all point-like 
interactions for which $\ri \gamma_t \gamma_x \prt_x$ extends 
to a self-adjoint Hamiltonian to the whole $\Gamma$. Observing that both the 
equation of motion (\ref{eqm2}) and the boundary condition (\ref{bc2}) preserve scale invariance, 
it is not surprising that the scattering matrix corresponding to these interactions is simply (\ref{crit1}). 

The Dirac field $\psi$ is complex, has a relativistic dispersion relation 
\begin{equation}
\omega(k) = |k| \, , 
\label{disp2}
\end{equation} 
and describes therefore both particle and antiparticle excitations. For quantizing 
(\ref{eqm2}, \ref{bc2}) we need for this reason two copies of 
reflection-transmission algebras \cite{bbm-un}. The first one $\A_+$ 
is generated by $\{a_i(k),\, a^*_i(k)\, :\, k \in \RR \}$ and $\S(k)$ given by (\ref{crit1}). We 
denote the second one by $\A_+^t$ because its generators 
$\{b_i(k),\, b^*_i(k)\, :\, k \in \RR \}$ obey the anti-commutation relations (\ref{rta1}, \ref{rta2}) 
with the {\it transpose} scattering matrix $\S^t(k)$. Besides (\ref{constr1}), one has therefore 
\begin{equation} 
b_i(k) = \sum_{j=1}^n b_j (-k) \S_{ji} (k) \, , \qquad 
b^\ast_i (k) = \sum_{j=1}^n \S_{ij} (-k) b^\ast_ j(-k) \, .    
\label{constr11}
\end{equation} 
In what follows we use the convention according to which $\{a_i(k),\, a^*_i(k)\}$ and 
$\{b_i(k),\, b^*_i(k)\}$ annihilate/create respectively antiparticles and particles. 
The solution of (\ref{eqm2},\ref{bc2}) in this basis is 
\begin{equation}
\psi_1(t,x,i) = 
\int_0^{\infty} \frac{\rd k}{2\pi} \left [a_i(k) \e^{-\ri k (t-x)} + b_i^*(-k) \e^{\ri k (t-x)} \right ]\, ,  
\label{psi21} 
\end{equation}
\begin{equation}
\psi_2(t,x,i) = \int_0^{\infty} \frac{\rd k}{2\pi} \left [a_i(-k) \e^{-\ri k (t+x)} + b_i^*(k) \e^{\ri k (t+x)}
\right ] \, . 
\label{psi22}
\end{equation} 

In the Dirac junction the anti-unitary operator $T$ of time reversal and the unitary operator 
$C$ of charge conjugation act as follows: 
\begin{equation}
T \psi_1(t,x,i) T^{-1} = \eta_T \psi_2(-t,x,i)\, , \qquad  
T \psi_2(t,x,i) T^{-1} = \eta_T \psi_1(-t,x,i)\, , \qquad |\eta_T|=1\, ,
\label{dtr}
\end{equation}
\begin{equation}
C \psi_1(t,x,i) C^{-1} = -\eta_C \psi_1^*(t,x,i)\, , \qquad  
C \psi_2(t,x,i) C^{-1} = \eta_C \psi_2^*(t,x,i)\, , \qquad |\eta_C|=1\, .
\label{cc1}
\end{equation}
Like in the Schr\"odinger case the boundary condition (\ref{bc2}) is invariant under time reversal 
only if $\UU$ is symmetric (\ref{tr2}). The condition for charge conjugation invariance is instead 
\begin{equation}
\overline \UU = -\UU\, . 
\label{cc2}
\end{equation}
The violation of (\ref{tr2}) and/or (\ref{cc2}) leads to the breakdown of the corresponding 
symmetry by means of the boundary condition (\ref{bc2}). 

The electric current and energy-momentum tensor are 
\begin{eqnarray} 
j_t(t,x,i) &=& -e :\psi^* \psi : (t,x,i) \, , \
\label{cdens}\\
j_x(t,x,i) &=& -e :\psi^* \sigma \psi : (t,x,i) \, ,  
\label{fcurr}
\end{eqnarray} 
\begin{eqnarray} 
\theta_{tt}(t,x,i) &=& \frac{\ri}{2}:\left [\psi^* (\prt_t \psi) - (\prt_t \psi^*) \psi \right ] : (t,x,i) \, , \qquad 
\label{endens}\\
\theta_{xt}(t,x,i) &=&  \frac{\ri}{2}:\left [\psi^* \sigma (\prt_t \psi) - (\prt_t \psi^*)\sigma \psi \right ] : (t,x,i) \, ,  
\label{encurr}
\end{eqnarray} 
where $:\cdots :$ denotes the normal product in $\A_+$ and $A_+^t$ and 
\begin{equation}
\sigma = \left(\begin{array}{cc}1 & 0\\ 0 & -1 \\ \end{array} \right)\, . 
\label{pauli}
\end{equation}

According to our convention for particles and antiparticles,  
the incoming asymptotic sub-algebra $\D_+^\inc$ is generated by 
$\{a_i(k),\, a^*_i(k), \,b_i(-k),\, b^*_i(-k)\, :\, k >0 \}$. The edge Hamiltonians 
and the asymptotic charge operators of particles and antiparticles are 
\begin{equation} 
h_i = \int_{-\infty}^0 \frac{\rd k}{2\pi} |k| \left [a^*_i(-k) a_i(-k) + b^*_i(k) b_i(k) \right ]\, , 
\label{ahd}
\end{equation} 
\begin{equation} 
q_i = \int_{-\infty}^0 \frac{\rd k}{2\pi} b^*_i(k) b_i(k) \, ,  \qquad 
\qt_i = -\int_{-\infty}^0 \frac{\rd k}{2\pi} a^*_i(-k) a_i(-k)\, , 
\label{acf1} 
\end{equation} 
respectively. We associate with (\ref{acf1}) the chemical 
potentials $\mu_i$ and $\tmu_i$. Now, following the general 
strategy explained in section 2, we set 
\begin{equation} 
K = \sum_{i=1}^n \beta_i (h_i -\mu_i q_i - \tmu_i \qt_i)  
\label{kopd}
\end{equation} 
and define first the steady state $\Omega_{\beta, \mu, \tmu}$ on the sub-algebra $\D_+^\inc$ 
by means of (\ref{def1}). Employing (\ref{constr1},\ref{constr11}), 
we extend after that the state $\Omega_{\beta, \mu, \tmu}$ to the whole algebra generated 
by $\A_+$ and $\A_+^t$. In this way one gets 
\begin{eqnarray} 
\langle a_j^*(p)a_i(k)\rangle_{\beta, \mu, \tmu} = 
2\pi \Bigl\{\Bigl [\theta(k)\ft_i(k) \delta_{ij}+ 
\theta(-k)\sum_{l=1}^n \UU^*_{il}\, \ft_l(-k)\, \UU_{lj}\Bigr ] \delta (k-p)  
\nonumber \\
+ \Bigl [\theta(k)\ft_i(k) \UU_{ij} + \theta(-k)\UU^*_{ij} \ft_j(-k) \Bigr ] \delta (k+p) \Bigr\}\, ,
\quad \; \; \,  
\label{cord1}
\end{eqnarray} 
\begin{eqnarray} 
\langle b_j^*(p)b_i(k)\rangle_{\beta, \mu, \tmu} = 
2\pi \Bigl\{\Bigl [\theta(-k)f_i(k) \delta_{ij}+ 
\theta(k)\sum_{l=1}^n \UU^*_{jl}\, f_l(-k)\, \UU_{li}\Bigr ] \delta (k-p)  
\nonumber \\
+ \Bigl [\theta(-k)\UU_{ji}^*f_i(k)  + \theta(k)f_j(-k)\UU_{ji}  \Bigr ] \delta (k+p) \Bigr\}\, ,
\quad \; \; \,  
\label{cord2}
\end{eqnarray} 
where 
\begin{equation}
\ft_i(k) = \frac{\e^{-\beta_i(|k| +\tmu_i)}}{1+\e^{-\beta_i(|k| +\tmu_i)}} \, , \qquad \quad 
f_i(k) = \frac{\e^{-\beta_i(|k| -\mu_i)}}{1+\e^{-\beta_i(|k| -\mu_i)}}  
\label{fgd}
\end{equation} 
are the Dirac distributions for antiparticles and particles respectively. 

Let us discuss finally the behavior of $\Omega_{\beta, \mu, \tmu}$ under charge conjugation. 
One easily verifies that $\Omega_{\beta, \mu, \tmu}$ is invariant under charge conjugation, namely 
$C \Omega_{\beta, \mu, \tmu}= \Omega_{\beta, \mu, \tmu}$, provided that both conditions 
\begin{equation}
\mu_i = -\tmu_i 
\label{cc3}
\end{equation}
and (\ref{cc2}) hold.  
\bigskip

\subsection{Transport properties} 
\medskip 

We are ready at this point to derive the steady currents in the state $\Omega_{\beta, \mu, \tmu}$. 
A computation, analogous to that performed in section 3, gives 
\begin{eqnarray}
J_i(\beta, \mu, \tmu ) \equiv \langle j_x(t,x,i) \rangle_{\beta, \mu, \tmu}  = 
e \sum_{j=1}^n \left (\delta_{ij} - |\UU_{ij}|^2 \right ) 
\int_0^{\infty} \frac{\rd k}{2\pi} \left [f_j(-k)-\ft_j(k)\right ]  
= \nonumber \\ 
\frac{e}{2\pi} \sum_{j=1}^n \left (\delta_{ij} - |\UU_{ij}|^2 \right ) \frac{1}{\beta_j} 
\ln \left ( \frac{1+\e^{\beta_j\mu_j}}{1+ \e^{-\beta_j \tmu_j}} \right )\, , 
\qquad \qquad \qquad \qquad \quad
\label{ff1}
\end{eqnarray} 
which satisfies Kirchhoff's rule and vanishes at equilibrium exactly like the Schr\"odinger steady current. 
A new feature of the Dirac steady current is the presence of particle and antiparticle contributions, captured 
respectively by the numerator and the denominator in the fraction under the logarithm in (\ref{ff1}). 

Concerning the dependence of (\ref{ff1}) on the chemical potentials, 
some particular cases of are worth mentioning. We first observe 
that if charge conjugation is preserved, the steady current (\ref{ff1}) vanishes because of (\ref{cc3}). 
This is due to a cancellation between the particle and antiparticle contributions. 
If instead all $\tmu_i = \mu_i$, the current (\ref{ff1}) is temperature independent (in spite 
of the fact that the junction is in contact with heat reservoirs with different temperatures $\beta_j$)  
and takes the simple form 
\begin{equation}
J_i(\beta, \mu, \mu ) = \frac{e}{2\pi}  \sum_{j=1}^n \left (\delta_{ij} - |\UU_{ij}|^2 \right ) \mu_j  \, . 
\label{ff2}
\end{equation} 
In the case $\tmu_i=0$ the steady current (\ref{ff1}) coincides with that of the Schr\"odinger junction (\ref{f2}) 
with scale invariant boundary conditions. 

Let us focus now on the temperature dependence of (\ref{ff1}). At high temperatures one has 
\begin{equation} 
\lim_{\beta_k=\beta \to 0} J_i(\beta, \mu, \tmu ) = 
\frac{e}{4\pi} \sum_{j=1}^n \left (\delta_{ij} - |\UU_{ij}|^2 \right ) 
\left (\mu_j  + \tmu_j \right )  \, , 
\label{ff3}
\end{equation} 
whereas at zero temperature
\begin{equation} 
\lim_{\beta_k=\beta \to +\infty} J_i(\beta, \mu, \tmu ) = 
\frac{e}{2\pi}  \sum_{j=1}^n \left (\delta_{ij} - |\UU_{ij}|^2 \right ) 
\left [\mu_j \theta (\mu_j) + \tmu_j \theta (-\tmu_j )\right ]  \, . 
\label{ff4}
\end{equation} 
The conductance tensor corresponding to (\ref{ff1}) is 
\begin{equation} 
\G_{ij}(\beta, \mu ) = \frac{e^2}{2\pi} \left (\delta_{ij} - |\UU_{ij}|^2 \right ) \frac{1}{\beta_j \mu_j} 
\ln \left ( \frac{1+\e^{\beta_j \mu_j}}{1+ \e^{-\beta_j\tmu_j}} \right )  
\label{cond22}
\end{equation}
and is not symmetric in general. 

For the energy flow one obtains 
\begin{eqnarray}
{\cal T}_i(\beta, \mu, \tmu ) = 
\langle \theta_{xt}(t,x,i)\rangle_{\beta, \mu, \tmu} = 
\sum_{j=1}^n \left (\delta_{ij} - |\UU_{ij}|^2 \right ) 
\int_0^{\infty} \frac{\rd k}{2\pi} \left [\ft_j(k)+f_j(-k)\right ]= \nonumber \\
\frac{1}{2\pi} \sum_{j=1}^n \left (|\UU_{ij}|^2 - \delta_{ij} \right ) \frac{1}{\beta_j^2} 
\left [{\rm Li}_2(-\e^{\beta_j \mu_j}) + {\rm Li}_2(-\e^{-\beta_j \tmu_j}) \right ]\, . 
\qquad \qquad \qquad 
\label{ff9}
\end{eqnarray} 

One can easily verify that the charge and energy densities are obtained by the replacement 
$\left (|\UU_{ij}|^2 - \delta_{ij} \right )\longmapsto \left (|\UU_{ij}|^2 + \delta_{ij} \right )$ in 
(\ref{ff1}) and (\ref{ff9}) respectively. Notice that these quantities are $x$-independent 
and therefore do not present Friedel oscillations. The reason is that both the dynamics (\ref{eqm2}) 
and the boundary conditions (\ref{bc2}) are scale invariant for massless fermions.  
\bigskip 

\subsection{Noise} 
\medskip 

{}For conciseness we report directly the zero-frequency noise power in terms of the matrix $\UU$ 
appearing in the boundary condition (\ref{bc2}) and the distributions (\ref{fgd}). One has 
\begin{eqnarray}
P_{ij}(\beta, \mu, \tmu) = e^2 \int^{\infty}_0 \frac{\rd k}{2 \pi}  
\Bigl \{ \delta_{ij} F_{ii}(k)  -|\UU_{ij}|^2 F_{ii}(k)  - |\UU_{ji}|^2 F_{jj}(k)  + \qquad \quad 
\nonumber \\
+\frac{1}{2}\sum_{l,m=1}^n \UU_{li}\overline{\UU}_{lj}  
\UU_{mj}\overline{\UU}_{mi} [F_{lm}(k)+F_{ml}(k)]\Bigr \}\, , 
\qquad \qquad \qquad 
\label{ff10}
\end{eqnarray}
with 
\begin{equation}
F_{ij}(k) =  f_{i}(k) [1-f_{j}(k)] + \ft_{i}(k) [1-\ft_{j}(k)] \, . 
\label{ff11}
\end{equation}
If all the temperatures are equal ($\beta_i = \beta$), the $k$-integration in the right 
hand side of (\ref{ff10}) can be performed exactly and gives 
\begin{eqnarray}
P_{ij}(\beta, \mu, \tmu) = \frac{e^2}{2 \pi \beta}  
\Bigl \{ (\delta_{ij} - |\UU_{ij}|^2)\left [ \frac{\e^{\beta \mu_i}}{1+\e^{\beta \mu_i}}
+  \frac{\e^{-\beta \tmu_i}}{1+\e^{-\beta \tmu_i}}\right ]
- |\UU_{ji}|^2 \left [ \frac{\e^{\beta \mu_j}}{1+\e^{\beta \mu_j}}
+  \frac{\e^{-\beta \tmu_j}}{1+\e^{-\beta \tmu_j}}\right ]  + \quad 
\nonumber \\
+\frac{1}{2}\sum_{l,m=1}^n \UU_{li}\overline{\UU}_{lj} \UU_{mj}\overline{\UU}_{mi} 
\left [\frac{\e^{\beta \mu_l} + \e^{\beta \mu_m}}{\e^{\beta \mu_l} - \e^{\beta \mu_m}} 
\ln \left (\frac{1+\e^{\beta \mu_l}}{1+\e^{\beta \mu_m}} \right ) + 
\frac{\e^{-\beta \tmu_l} + \e^{-\beta \tmu_m}}{\e^{-\beta \tmu_l} - \e^{-\beta \tmu_m}} 
\ln \left (\frac{1+\e^{-\beta \tmu_l}}{1+\e^{-\beta \tmu_m}} \right )
\right ]\Bigr \}. \qquad 
\label{ff12}
\end{eqnarray}
{}For the purely thermal noise on gets therefore 
\begin{equation}
P_{ij}(\beta, 0, 0) = \frac{e^2}{2\pi \beta}  
\left (2\delta_{ij} - |\UU_{ij}|^2 - |\UU_{ji}|^2 
\right ) \, , 
\label{ff13}
\end{equation}
which coincides precisely with the result (\ref{N10}) for the Schr\"odinger junction at criticality. 

\bigskip 

\subsection{Remarks} 
\medskip 

In spite of the different dispersion relations, 
the general structure of the steady currents and the noise in the Schr\"odinger and Dirac cases are quite similar. 
A characteristic feature of the Dirac case is the possibility to introduce the independent chemical 
potentials $\mu_i$ and $\tmu_i$, associated with particles and antiparticles. 
This fact has elementary but important consequences. If $\mu_i=-\tmu_i$ 
the particle and antiparticle contributions cancel each other in the electric steady current (\ref{ff1}), but sum up in 
the heat current (\ref{ff9}) and in the zero frequency noise (\ref{ff13}). 

{}Following the argument in section 3.5, the 
results about the Dirac junction have a straightforward generalization to the case when $\psi$ is 
minimally coupled to a static classical electromagnetic field generated by the potential $(A_t(P), \abf (P))$ 
in the ambient space. 
\bigskip

\section{Outlook and conclusions} 
\medskip

In this paper we developed an algebraic method for constructing non-equilibrium steady states 
$\Omega_{\beta, \mu}$ on star graphs. Our approach is microscopic and our construction 
generalizes that of a Gibbs state over the algebra of canonical (anti)commutation relations. 
The Schr\"odinger and Dirac equations have been investigated in this framework. We considered in detail 
the case in which the interaction, driving the system away from equilibrium, is localized in the 
vertex of the graph. It turns out that the non-equilibrium dynamics, generated by such interactions, 
is exactly solvable. In fact, the $\Omega_{\beta, \mu}$-expectation values of various observables 
(currents and charge densities) can be computed exactly, without resorting to any kind of 
approximation. We have shown in particular, that the 
expectation value of the electric current in the Schr\"odinger case 
reproduces precisely the famous L-B formula. Once the formalism has been tested on the L-B steady 
current, we applied it for the computation of the charge and energy densities, the energy flow and the noise power. 
The presence of Friedel oscillations has been detected. We demonstrated also that 
point-like interaction in the junction modifies the linear dependence of the thermal noise 
on the temperature (Johnson-Nyquist formula). The formalism has been generalized in order to include 
the minimal coupling to an external time-independent electromagnetic field as well. 

Summarizing, the star graph models proposed and analyzed in this paper 
represent relatively simple exactly solvable examples of quantum non-equilibrium systems 
in a steady state. For this reason they provide a nice laboratory for testing 
general ideas about non-equilibrium dynamics. 

Our results can be generalized in various directions. 
First of all, one can consider more complicated networks 
with several junctions and loops, which can be crossed by magnetic fluxes. The basic idea for treating this case 
is to replace in the above formalism the scattering matrix $\S$ with an effective one 
$\S_{\rm eff}$, which takes into account all vertex interactions  \cite{ks-01}-\cite{Khachatryan:2009xg} and the 
presence of a magnetic field (see (\ref{a8})). The derivation of the L-B steady current and the noise in this case is  
of particular physical interest and is currently under investigation \cite{cmr-11}. 

Another possible generalization is the study of imperfect leads involving interactions with external potentials 
and/or self-interactions like those in the Luttinger liquid. More general boundary interactions, involving new 
vertex degrees of freedom of the type appearing in the resonant-level model \cite{bd-11}, can be investigated in the 
above framework as well. We will discuss these issues elsewhere. 

\bigskip 
\leftline{\bf Acknowledgments:} 
\medskip 

It is a great pleasure to thank B. Bellazzini, V. Caudrelier, P. Sorba and E. Ragoucy for the intense collaboration on 
quantum field theory on graphs. Enlightening discussions and correspondence 
with P. Calabrese, B. Dou\c{c}ot, V. Kostrykin, I. Safi, R. Schrader and E. Vicari are also kindly acknowledged.



\begin{thebibliography}{99} 


\bibitem{kf-92} 
C.~L.~Kane and M.~P.~A. Fisher, Phys. Rev. Lett. {\bf 68}, 1220 (1992); 
Phys. Rev. {\bf B 46}, 15233 (1992). 

\bibitem{SS}
I.~Safi, H.~J.~Schulz, 
Phys.\ Rev.\  B {\bf 52}, R17040 (1995).

\bibitem{nfll-99} 
C. Nayak, M. P. A. Fisher, A.~W.~W.~Ludwig and H.~H.~Lin, 
Phys.\ Rev.\  B {\bf 59}, 15694 (1999).

\bibitem{sdm-01} I. Safi, P. Devillard, and T. Martin, 
Phys. Rev. Lett. {\bf 86}, 4628 (2001). 

\bibitem{mw-02} J.E. Moore and X.-G. Wen, Phys. Rev. B {\bf 66}, 115305 (2002).

\bibitem{y-02} H. Yi, Phys. Rev. B {\bf 65}, 195101 (2002).

\bibitem{lrs-02} 
S. Lal, S. Rao, and D. Sen, Phys. Rev. B {\bf 66}, 165327 (2002).

\bibitem{cte-02} 
S. Chen, B. Trauzettel, and R. Egger, 
Phys. Rev. Lett. {\bf  89}, 226404 (2002).

\bibitem{ppil-03} K-V. Pham, F. Piechon, K-I Imura, P. Lederer,
Phys. Rev. B {\bf 68}, 205110 (2003).

\bibitem{coa-03} C. Chamon, M. Oshikawa, and I. Affleck,
Phys. Rev. Lett. {\bf 91}, 206403 (2003); 
M. Oshikawa,  C. Chamon, and I. Affleck,  J. Stat. Mech. P02008 (2006). 

\bibitem{dgst-03} F. Dolcini, H. Grabert, I. Safi and B. Trauzettel, 
Phys. Rev. Lett. {\bf 91}, 266402 (2003). 

\bibitem{rs-04} S. Rao and D. Sen, Phys. Rev. B {\bf 70}, 195115 (2004). 

\bibitem{kd-05} K.~Kazymyrenko and B.~Dou\c{c}ot,
Phys. Rev. B {\bf 71}, 075110 (2005)

\bibitem{klvf-05}
E.-A. Kim, M. J. Lawler, S. Vishveshwara, E. Fradkin, 
Phys. Rev. Lett. 95, 176402 (2005); 
Phys. Rev. B 74, 155324 (2006). 

\bibitem{emabms-05} T.~Enss, V.~Meden, S.~Andergassen, X.~Barnabe-Theriault,
W.~Metzner, K.~Schonhammer, Phys. Rev. B {\bf 71}, 155401 (2005);
X.~Barnabe-Theriault, A~ Sedeki, V.~Meden, K.~Schonhammer, 
Phys. Rev. Lett. {\bf 94}, 136405 (2005);
X.~Barnabe-Theriault, A.~Sedeki, V.~Meden, K.~Schonhammer, 
Phys. Rev. B {\bf 71}, 205327 (2005). 

\bibitem{ff-05}
D. Friedan, cond-mat/0505084; cond-mat/0505085.

\bibitem{drs-06} S.~Das, S.~Rao, D.~Sen, Phys. Rev. B {\bf 74}, 045322 (2006).

\bibitem{Bellazzini:2006jb} 
B.~Bellazzini and M.~Mintchev, 
J.\ Phys.\ A  {\bf 39}, 11101 (2006) 
[arXiv:hep-th/0605036]. 

\bibitem{Bellazzini:2006kh} 
B.~Bellazzini, M.~Mintchev and P.~Sorba,
J.\ Phys.\ A  {\bf 40}, 2485 (2007)
[arXiv:hep-th/0611090].

\bibitem{Bellazzini:2008mn} 
B.~Bellazzini, M.~Burrello, M.~Mintchev and P.~Sorba,
Proc. Symp. Pure Math. {\bf 77}, 639 (2008), 
arXiv:0801.2852 [hep-th].

\bibitem{Bellazzini:2008fu} 
B.~Bellazzini, P.~Calabrese and M.~Mintchev, 
Phys.\ Rev.\  B {\bf 79} 085122 (2009), 
[arXiv:0808.2719]. 

\bibitem{hc-08}
C.-Y. Hou and C. Chamon, Phys. Rev. B {\bf 77}, 155422 (2008).

\bibitem{drs-08}
S. Das, S. Rao, and A. Saha, Phys. Rev. B {\bf 77}, 155418 (2008);
Europhys. Lett. {\bf 81}, 67001 (2008). 

\bibitem{dr-08} S. Das and S. Rao, Phys. Rev. B {\bf 78}, 205421 (2008). 

\bibitem{hkc-08}
C.-Y. Hou, E.-A. Kim, and C. Chamon, Phys. Rev. Lett. {\bf 102}, 076602 (2009).  

\bibitem{adrs-08} 
A. Agarwal, S. Das, S. Rao and D. Sen, 
Phys. Rev. Lett. {\bf 103}, 026401 (2009);   
Erratum, Phys. Rev. Lett. {\bf 103}, 079903 (2009), 
arXiv:0810.3513 [cond-mat]. 

\bibitem{dr-09} 
S. Das, S. Rao and A. Saha, Phys. Rev. B {\bf 79}, 155416 (2009). 

\bibitem{Bellazzini:2008cs} 
B.~Bellazzini, M.~Mintchev and P.~Sorba, 
J. Math. Phys. {\bf 51} 032302 (2010), 
arXiv:0810.3101 [hep-th].

\bibitem{Ines}
I.~Safi, 
arXiv:0906.2363 [cond-mat.] 

\bibitem{Bellazzini:2009nk} 
B.~Bellazzini, M.~Mintchev and P.~Sorba, 
Phys.\ Rev.\  B {\bf 80} 25441 (2009), 
arXiv:0907.4221[hep-th].

\bibitem{Bellazzini:2010gs}
B.~Bellazzini, M.~Mintchev and P.~Sorba, 
Phys.\ Rev.\  B {\bf 82} 195113 (2010), 
arXiv:1002.0206 [hep-th].

\bibitem{Soori:2010ga}
A.~Soori, D.~Sen, 
Europhys.\ Lett.\  {\bf 93 } (2011) 57007.


\bibitem{mcl-59}
J. A. McLennan, 
Phys. Rev. {\bf 115} (1959) 1405. 

\bibitem{els-96}
G. L. Eyink, J. L. Lebowitz and H. Spohn, 
J. Stat. Phys. {\bf 83} (1996) 385. 

\bibitem{ru-00} 
D. Ruelle, 
J. Stat. Phys. {\bf 98} (2000) 57. 

\bibitem{jl-01} 
L. Bertini, A. D. Sole, D. Gabrielli, G. Jona-Lasinio and C. Landim, 
Phys. Rev. Lett. {\bf 87} (2001) 040601. 

\bibitem{jp-02} 
V. Jaksic and C. A. Pillet,  
J. Stat. Phys. {\bf 108} (2002) 787. 

\bibitem{bd-04}
T. Bodineau and B. Derrida, 
Phys. Rev. Lett. {\bf 92} (2004) 180601. 

\bibitem{st-06}
S. Sasa and H. Tasaki, 
J. Stat. Phys. {\bf 125} (2006) 125. 

\bibitem{d-10}
B. Derrida, arXiv:1012.1136 [cond-mat.stat-mech]


\bibitem{ks-00}
V.~Kostrykin and R.~Schrader, 
Fortschr. Phys. {\bf 48}, 703 (2000). 

\bibitem{h-00}
M.~Harmer, 
J.\ Phys.\ A {\bf 33} (2000) 9015. 

\bibitem{k-08}
P. Kuchment, arXiv:0802.3442 [math-ph]. 


\bibitem{la-57}
R. Landauer, 
IBM J. Res. Dev. {\bf 1} (1957) 233;  
Philos. Mag. {\bf 21} (1970) 863.

\bibitem{bu-86}
M. B\"uttiker, 
Phys. Rev. Lett. {\bf 57} (1986) 1761; 
IBM J. Res. Dev. {\bf 32} (1988) 317. 


\bibitem{KTH} R.~Kubo, M.~Toda and N.~Hashitsume, {\it Statistical Physics II}, (Springer, Berlin, 1985). 


\bibitem{bs-89} 
H.~U.~Baranger and A.~D.~Stone, 
Phys. Rev. B {\bf 40} (1989) 8169. 

\bibitem{cj-05} 
H.~D.~Cornean and A.~Jensen, 
J. Math. Phys. {\bf 46} (2005) 042106-1. 


\bibitem{Sch} 
R.~Schrader, 
J. Phys. A {\bf 42} (2009) 495401, 
arXiv:0907.1522 [hep-th]. 


\bibitem{Liguori:1996xr} 
A.~Liguori, M.~Mintchev and L.~Zhao, 
Commun.\ Math.\ Phys.\  {\bf 194}, 569 (1998)
[arViv:hep-th/9607085]. 

\bibitem{Mintchev:2002zd} 
M.~Mintchev, E.~Ragoucy and P.~Sorba, 
Phys.\ Lett.\  B {\bf 547}, 313 (2002) 
[arXiv:hep-th/0209052]. 

\bibitem{Mintchev:2003ue} 
M.~Mintchev, E.~Ragoucy and P.~Sorba, 
J.\ Phys.\ A  {\bf 36}, 10407 (2003) 
[arXiv:hep-th/0303187]. 

\bibitem{Mintchev:2004jy} 
M.~Mintchev and P.~Sorba, 
J.\ Stat.\ Mech.\  {\bf 0407}, P001 (2004) 
[arXiv:hep-th/0405264]. 



\bibitem{BR} O.~Bratteli and D.~W.~Robinson, {\it Operator Algebras and Quantum Statistical 
Mechanics 2}, (Springer, Berlin, 1996). 


\bibitem{fri-52} J.~Friedel, 
Phil. Mag. {\bf 43} (1952) 153. 


 
 \bibitem{il-99} 
 Y.~Imry and R.~Landauer, 
 Rev. Mod. Phys. {\bf 71} (1999) S306. 
 
 
 \bibitem{mla-92}
 Th.~Martin and R.~Landauer, 
 Phys. Rev. B {\bf 45} (1992) 1742. 
 
 \bibitem{bu-92}
 M.~B\"uttiker, 
 Phys. Rev. B {\bf 46} (1992) 12485. 
 
 
 \bibitem{bb-00} 
 Ya.~M.~Blanter and M.~B\"uttiker, 
 Phys. Rep. {\bf 336} (2000) 1. 
 

 
 
 \bibitem{gr-07}
 I.~S.~Gradshteyn and I.~M.~Ryzhik, {\it Tables of Integrals, Series and Products}, (Elsevier, Amsterdam, 2007). 
 
 \bibitem{ksch-03}
V.~Kostrykin and R.~Schrader, 
Commun. Math. Phys. {\bf 237} (2003) 161. 


\bibitem{bh-03}
J. Bolte and J. Harrison, 
J.\ Phys.\ A  {\bf 36} (2003) 2747. 

 
\bibitem{bbm-un} 
B.~Bellazzini, M.~Burrello and M.~Mintchev, unpublished. 


\bibitem{ks-01} 
V.~Kostrykin and R.~Schrader, 
J. Math. Phys. {\bf 42} (2001) 1563. 

\bibitem{Mintchev:2007qt} 
M.~Mintchev and E.~Ragoucy, 
J.\ Phys.\ A  {\bf 40} (2007) 9515 
[arXiv:0705.1322 [hep-th]]. 

\bibitem{Ragoucy:2009hf}
E.~Ragoucy, 
J.\ Phys.\ A  {\bf 42} (2009) 295205, 
[arXiv:0901.2431 [hep-th]]. 

\bibitem{Caudrelier:2009ay} 
V.~Caudrelier and E.~Ragoucy, 
Nucl.\ Phys.\  B {\bf 828} (2010) 515, 
[arXiv:0907.5359 [math-ph]]. 

\bibitem{Khachatryan:2009xg} 
S.~Khachatryan, A.~Sedrakyan and P.~Sorba, 
Nucl.\ Phys.\  B {\bf 825} (2010) 444, 
[arXiv:0904.2688 [cond-mat.mes-hall]].

\bibitem{cmr-11} 
V.~Caudrelier, M.~Mintchev and E.~Ragoucy, (work in progress). 

\bibitem{bd-11} 
D.~Bernard and B. Doyon, 
arXiv:1105.1695 [math-ph]

 
 
 
 
\end{thebibliography}
\end{document}